# Surface melting of nanoscopic epitaxial films

**P. Müller[1], R. Kern**
Centre de Recherche sur les Mécanismes de la Croissance Cristalline[2]
Campus de Luminy, case 913, F-13288 Marseille Cedex 9, France

## ABSTRACT

*By introducing finite size surface and interfacial excess quantities, interactions between interfaces are shown to modify the usual surface premelting phenomenon. It is the case of surface melting of a thin solid film s deposited on a planar solid substrate S. More precisely to the usual wetting condition of the solid s by its own melt l, necessary for premelting (wetting factor Φ<0), is adjoined a new quantity Γ describing the interactions of the l/s interface with the s/S interface. When Γ>0 this interface attraction boosts the premelting so that a two stage boosted surface premelting is foreseen: a continuous premelting, up to roughly half the deposited film, is followed by an abrupt first order premelting. When Γ<0 these interfaces repell each other so that premelting is refrained and the film remains partly solid above the bulk melting point (overheating) what is called astride melting. Elastic stress modifies both types of melting curves. Bulk and surface stresses have to be distinguished.*

*For coherent epitaxial layers the natural misfit determining the strain and the elastic energy density (independent of the thickness of the solid) only shifts the melting curves to lower temperature, up to thicknesses where strain relief happens. Surface stress, as a finite size surface excess quantity, modifies the wetting factor Φ and the coefficient Γ, therefore the wetting properties and thus the melting curves are slightly modified. For perfect glissile epitaxies things are more complex since bulk strain and elastic energy density (now induced by surface stress) varies with the film thickness. The melting curves are thus distorted on their initial part (either in the sense of assisted or refrained premelting) depending upon the set of interfacial stresses.*

*Lastly there is a z-inhomogeneity of stress due to the interactions between the bulk of the various material layers. This leads to measurable strain gradients in the film but only distorts the final part of the melting curve.*

*Some of these theoretical results have been experimentally illustrated in the Γ<0 case where then useful interfacial data, adhesion energies and interfacial stress data have been collected but the Γ>0 case remains fully open to future exploration.*

---

[1] Email: muller@crmc2.univ-mrs.fr
[2] Associé aux Universités Aix-Marseille II et III.

## Introduction

It is now well-known that when a solid surface is wetted by its own melt, in equilibrium conditions, a liquid phase may cover this surface at a temperature below its bulk melting point $T_m$. As the increasing temperature of the solid approaches $T_m$ from below the thickness of the liquid film increases continuously and diverges asymptotically at $T=T_m$. Such a phenomenon has been called surface pre-melting or *surface induced melting* and has been thoroughly studied both from an experimental and a theoretical view-point (for reviews see [1-4]). The reverse phenomenon, the formation of crystalline layers on the surface of its liquid above the melting temperature $T_m$, has also been reported [5,6]. The generic term of *surface induced freezing* has thus been proposed to describe a continuous transition in which a more-ordered surface phase grows on a less-ordered bulk phase. In figure 1 are shown schematically the two cases of surface melting (1a) and surface freezing (fig 1b).

Concerning surface melting, most of the studies concern the surface pre-melting of semi-infinite solids [7,8] and interesting peculiarities have been discovered: **(i)** two successive asymptotic laws for approaching $T_m$ [9-11]; **(ii)** incomplete premelting (at $T<T_m$ premelting stops its progression up to $T_m$ where the usual first order melting takes place). **(iii)** incomplete wetting and premelting (at $T_w<T_m$ some wetting layers suddenly appear. Their thickness then asymptotically increases towards $T_m$). Both phenomena (ii) and (iii) received their theoretical framework by the concept of surface-induced layering in liquids [12,13] opposing to surface-induced disordering leading to the usual type of premelting. Distinction between surface melting and surface roughening has been clearly done by [14].

Only a few papers concern the case of finite size solids. *Due to the finite size some new effects should be expected and are discussed in this paper*. It is the case of the surface induced melting of nanometric solid particles for which furthermore the melting temperature of the bulk depends upon the size of the particle. In this case the number of liquid layers increases continuously with temperature until the core of the particle melts suddenly at its curvature-dependent melting point [15-17]. Finite size effects of the solid phase can also be put in evidence in the case of thin films. Bienfait, Dash and their collaborators [18-23] have studied by various methods the surface premelting of deposited adsorbed simple gases forming thin solid films with some strain. They speak of strain assisted premelting and infer that substrate interaction may retain solid layers near to the substrate interface. A model of melting of several solid layers has been proposed by Petterson et al [24] and then by [23] trying to formalise the various observed effects. Lastly H.Sakai [25] recently showed theoretically that a free thin slab may exhibit a two stage melting transition. In a first stage the equilibrium thickness of the premelted liquid continuously increases with temperature, then below some critical temperature $T_c < T_m$ and thickness $d_c < d$ of the solid the slab melts completely (first order transition). The transition temperature of the second stage was associated with the thickness of the slab.

The aim of this paper is to revisit the surface pre-melting of epitaxially deposited films. More precisely *we want to elucidate both the size effect and the strain effect* when the epitaxial layers are pseudomorphic (or not) to their substrate S (see fig 2a). For this purpose in section I we define our model with special care on the necessary assumptions. Let us underline that surface freezing (fig 2b) can be simply obtained from surface melting (fig 2a) by interchanging the liquid l and solid s. Therefore, all the following results on surface melting remain valid for surface freezing by interchanging in the formulae and diagrams the indices s and l and changing the latent melting entropy $\Delta S_m$ in $\Delta S_{freez} = -\Delta S_m$ the latent freezing entropy. However we do not concentrate on surface freezing which concerns in fact relatively exotic material as liquid crystals [5,6]. In section II we write the free enthalpy of such a system (fig 2a), then in II1 we seek for equilibrium and stability conditions, II2 *leading to two different new surface melting regimes*. In case of coherent epitaxies, both regimes are identically scaled in temperature by the strain energy. The two regimes are

(i)     continuous partial premelting relayed by discontinuous first order melting (boosted premelting)

(ii)    continuous premelting relayed by continuous overheating (astride melting)

In both cases the usual wetting condition of the solid s by its melt l is fulfilled but a new parameter determines whether the l/s interface is attracted (i) or repelled by the substrate S in what case (ii) the last solid layers resist to melting. For thick solid films the two melting regimes go over to the usual asymptotic surface melting.

In section III we discuss the two regimes numerically, look at their possible occurrence in III2, propose some possible experiments to measure independently adhesion energies and treat an example of well studied astride surface melting in III4. In section IV we introduce the notion of surface stress, we neglected in section II, and show that its consideration may be a valid correction for coherent epitaxies. Its contribution is however crucial in the case of incoherent glissile epitaxy we treat in section V where an experimental example is given. In VI we consider how the epitaxial layer becomes inhomogeneously strained by the substrate field and discuss how it acts on the foregoing effects. Finally in section VII we conclude and give an outlook about what would be of interest to investigate by experiments in view of our predictions.

## I/ Model of surface-induced melting of pseudomorphous films

For that purpose we consider a semi-infinite planar substrate S of material B supposed to be chemically inert in respect to the deposit A. The melting point of S ($T_S$) is much higher that the melting point $T_m$ of A so that experiments can be done around $T_m$ without alteration of B and A.

Material A is either elemental or a defined compound with congruent melting so that the solid s has the same composition as its liquid l and has a defined melting point $T_m$.

Substrate S of material B bears a lattice-mismatched composite material A s+l of $n_s$ solid layers and $n_l$ liquid layers (fig.2a). The $n_s$ layers are in pseudomorphous contact and epitaxially stressed by S whereas its $n_l$ upper layers A are in the liquid state. For the sake of simplicity, materials A and B are supposed to be cubic of respective parameters a and b. Their surfaces (001) are in contact with parallel orientation of the in-plane axis a and b. The in-plane natural misfit therefore is m=(b-a)/a [3].

In the case of this coherent epitaxy we study in the following sections, the solid s is homogeneously in-plane strained by the amount $\varepsilon_{xx}=\varepsilon_{yy}=m$ with m=(b-a)/a the natural misfit of the contact which becomes equal to the in-plane strain when A(001) is rendered coherent with B(001). In VI inhomogeneous strained layers are considered. Partial strain relaxation or loss of coherence is only considered as event appearing at some greater thickness disrupting the process we describe. The case of non coherent but glissile epitaxies is considered in V.

The (001) bare surface of A and B as well as the various interfaces are supposed to be morphologically stable. They do not suffer facetting. Therefore these surfaces and interfaces have in their own orientation an inward cusp in their γ-plot. These faces are usually so-called F or singular faces. In summary the system, even when strained, remains planar at microscopic and mesoscopic scale during a raise of temperature up to $T_m$.

The last but essential point is that our model of surface melting uses the notion of finite size surface and interfacial specific energies. It is known that Landau's theory of phase transitions has predicted surface-melting [27-28] in very general terms of order parameter. However

---

[3] Epitaxy is however not limited to the regular overgrowth of cubic species with parallel axis. Its realm is much more rich and its crystallographic rules less degenerated: two species whatever their chemical nature and symmetries may gather two lattice planes and in theses planes one or two pairs of lattice rows. See for a review [26]

relations of quantitative interest could only be obtained by adjoining models. A two parabola model [8] leads to temperature dependant order parameter profiles and finally to an explicit minimal surface free energy. Pluis et al [8] could identify the model parameters so that the surface free energy of the system becomes thickness dependent. Two terms describe the creation of the film l of thickness h on the dry surface of s: the positive bulk melting free energy proportional to h vanishing at $T=T_m$ and the surface energy change of wetting $\Delta\gamma \exp(-2h/\zeta)$ with $\Delta\gamma = \gamma_s - \gamma_l - \gamma_{sl} > 0$ where $\zeta$ is a correlation length in the liquid but otherwise non precised. When $\Delta\gamma>0$, not only the melt wets its solid but both free energies oppose so that some equilibrium thickness $h_{eq}$ is installed. One can speak also [8] in terms of thermodynamic forces or effective forces opposing each other. This effective force interpretation can be directly extended to finite-size liquids and finite-size solids as in fig 2 so that now appear forces between the interfaces l/s-l/v; l/v-s/S and l/s-s/S. In appendix III this is done in details for some planar systems. Let us remark that this effective force interpretation in surface melting has been introduced in 1968 by Bolling [29,30] for grain boundary melting and in 1972 for the specific case of surface melting of ice by Lacmann and Stranski [31,32].

## II/ Free energy of the system

Our purpose being to seek for the equilibrium number of liquid layers as a function of temperature, we have to minimise a thermodynamic potential of the composite system constituted of liquid and solid layers of A sitting on a semi-infinite substrate S (see fig. 2a). We will note this system l/s/S. Since the $n_s$ layers are epitaxially strained by S, the liquid submitted eventually to hydrostatic pressure P, we have to use a good thermodynamic potential. In appendix I following [33] and [34] it is shown that the Gibbs free energy per unit area of S is pertinent for the solid s and the liquid l provided correct boundary stress-strain conditions are applied at the surface and interface. In appendix II are written the mechanical boundary conditions of the coherent epitaxial system. Therefore we have to minimise :

$$G=N_sG^s+N_lG^l+G^{surf} \qquad (1)$$

where $G^s$ is the Gibbs free energy per solid mole (number of moles $N_s$ per unit area), $G^l$ the Gibbs energy per liquid mole (number of moles $N_l$ per unit area) and $G^{surf}$ the excess energy due to surface and interfaces. The total number of moles of material A thus is

$$N=N_s+N_l \qquad (2)$$

In (1) $G^{surf}$ is the surface excess enthalpy of the system of fig 2a. In spite of its areal constancy it is size dependant since it is thickness dependant for nanoscopic thicknesses and then must read $G^{surf}(n_s,n_l)$ where $n_s$ and $n_l$ are the number of solid and liquid layers respectively, n being the total number of layers..

$$n=n_s+n_l \qquad (3)$$

Note that the relation between $N_i$ and $n_i$ is:

$$n_s=N_s\left(v^s\right)^{2/3}=N_s\left(v^s\right)(1+m)^2, \; n_l=N_l\left(v^l\right)^{2/3} \qquad (4)$$

where $v^i$ are the atomic volumes of s,l and S and m the natural misfit as defined in I. According to appendix I (formula III) the free enthalpy of the liquid (supposed to be under zero hydrostatic pressure P=0 since its vapour pressure around the melting point is negligible in respect to the epitaxial stress $\sigma_{11}$ by a factor of $10^{-5}$) reads:

$$G^l(T,0)=U^l(T,0)-TS^l(T,0) \qquad (5)$$

At the same that one of the coherent solid reads (see appendix I, formula II and appendix II, formula I):

$$G^s\left(T,\sigma_{ij}\right)=U^s(T,0)-TS^s(T,0)+V^s\,\frac{Em^2}{1-\nu} \qquad (6)$$

$U^i$ and $S^i$ are the molar internal energy and entropy of the solid and the liquid i=s,l at zero pressure, $V^s$ the molar volume of the solid at zero stress, E and $\nu$ the Young'modulus and Poisson' ratio of the solid s in the proper orientation (see footnote of appendix II). All quantities are defined at temperature T even the natural misfit m=(b-a)/a that may be sensibly temperature dependent via the in-plane differential dilatation of s and S.

We have now to write the surface free energy contribution $G^{surf}$ of (1). For the planar system (fig 2a) we consider, there are three interfaces. As stated at the end of section I, the corresponding free energies are thickness-dependent for thin film of l and s (size effect). For illustration let us take a thin slab of $n_i$ layers of body i. When creating such a slab by extraction from an infinite body of i (bringing both remaining semi-infinite bodies again together), the two created surfaces of the slab have a total surface free energy $2\gamma_i(n_i)$ function of $n_i$. When this slab becomes thick $n_i\rightarrow\infty$ this energy has to tend towards the macroscopic surface free energy one defines usually  by separation of an infinite body i in two semi-infinite bodies that is $2\gamma_i(\infty)$. When extracting *only a monolayer*, the other limiting value of surface energy is $2\gamma_i(1)$. The value $\gamma_i(1)$ is therefore also the work of binding a monolayer of unit area with all j=1,2,…. $\infty$ underlying layers of the semi-infinite body i. In the hypothetical case there are only first neighbours interactions with these layers, the binding work of a bilayer would be  $\gamma_i(2)=\gamma_i(1)$ since the second layer of the bilayer does not feel the substrate layers. More generally one has $\gamma_i(n_i)=\gamma(\infty)$, $(1\leq n_s<\infty)$ that means there is no size effect. However second, third and so on layers-interactions exist even if small and rapidly decreasing in condensed matter. A convenient formalisation is to write $\gamma_i(n_i)=\gamma_i(\infty)f(n_i)$ with $f(n_i)$ a continuous decreasing function with $f(\infty)=1$ and $0<f(1)<1$. The exponential dependence $f(n_i)=1-e^{-n_i/\zeta_i}$ is of that type (with $\zeta_i$ a characteristic of phase i) and may have some physical meaning. In appendix III we justify this analytical expression and give the free enthalpy per unit substrate area of the l/s/S system of figure 2a as a function of the number of layers $n_l$ and $n_s$.

$$G^{surf}(n_s,n_l)=N_{avo}\left[2\gamma_S+\left(2\gamma_l-\beta_{l/s}\right)\left(1-e^{-n_l/\zeta_l}\right)+\left(2\gamma_s-\beta_{s/S}\right)\left(1-e^{-n_s/\zeta_s}\right)+\left(\beta_{l/s}-\beta_{l/S}\right)e^{-n_s/\zeta_s}\left(1-e^{-n_l/\zeta_l}\right)\right]$$

(7)

$\gamma_i$ and $\beta_{i/j}$ are the surface and adhesion energies of i=l, s, S or i/j=l/s, l/S, s/S of the macroscopic phases. $\zeta_l$ and $\zeta_s$ give a measure of finiteness of the interactions of the liquid and of the solid, $N_{avo}$ the Avogadro number.

Let us remark here that according to (1), and (5) to (7) using the Gibbs procedure we divided G in bulk and surface excess quantities. The surface excess quantities (7) therefore bear the finite size effects and not the bulk quantities (5) and (6). It would be inconsistent to introduce for these bulk quantities some finite size properties.

Let us stress some limiting remarks: **(i)** we should remind that the solid part of the film was brought into coincidence with substrate S by uniaxial $\varepsilon_{11}=\varepsilon_{22}=m$ deformation but we don't change the surface energy term (7). Therefore we have to amend it. For easiness this will be done only in section IV concerning surface stress. **(ii)** we should mention too that due to the finiteness of the film, surface excess (7) implies that there is an excess potential inside the layers l and s so that they become elastically inhomogeneous. Again for easiness we will treat it only in section VI.

## II.1/ Equilibrium conditions:

The stationary number of liquid layers can be obtained by the condition $\partial G/\partial N_l\big|_N=0$ with (1) and (5) to (7). For essentialness, we use (4) by neglecting the size differences of $v^s$ and $v^l$ and suppose $\zeta_s=\zeta_l=\zeta$, thus $\partial G/\partial N_l\big|_{N,T}=0$ reads:

$$\Delta U_m(T)-T\Delta S_m(T)-\frac{V^s Em^2}{1-\nu}+N_{av}\frac{b^2}{\zeta}\left[\Phi e^{-n_l/\zeta}-\Gamma e^{-n_s/\zeta}\right]=0 \qquad \textbf{(8)}$$

where $\Delta U_m = U^l - U^s$, $\Delta S_m = S^l - S^s$ are the melting energy and entropy respectively. The quantities $\Phi$ and $\Gamma$ read:

$$\Phi = 2\gamma_l - \beta_{l/s} \equiv \gamma_l + \gamma_{ls} - \gamma_s \qquad (9)$$

$$\Gamma = (2\gamma_s - \beta_{s/s}) + (\beta_{l/s} - \beta_{l/s}) \equiv \gamma_{sS} + \gamma_{ls} - \gamma_{lS} \qquad (10)$$

where all specific energies are those of the macroscopic phases with planar surfaces $\gamma_i$, $\gamma_{ij}$ or $\beta_{i/j}$. They are slightly temperature dependent as $E, V^s$ are, so that we consider the value they take in the following near $T_m$. In (9) and (10) the first expressions are written in terms of surface and adhesion energies. The second expressions are obtained by using Dupré' equation (See III in appendix III) in terms of surface and interfacial energies. These factors $\Phi$ and $\Gamma$ will play an essential role in the further classification of the predicted phenomena.

Since at the bulk melting point $T_m$ (without stress) there is $\Delta U_m(T_m) = T_m \Delta S_m(T_m)$, neglecting the heat capacity change at constant pressure $\Delta C_m$ one has not too far from the melting point the linear dependence:

$$\Delta U_m(T) - T\Delta S_m(T) \approx \Delta S_m(T_m)(T_m - T) \qquad (11)$$

where $\Delta S_m (T_m)$ is the latent melting entropy at the melting point $T_m$ which according to Matignon' rule amounts to $\Delta S_m(T_m) = 2\text{-}3$ cal mole$^{-1}$ deg$^{-1}$ for most elements. In the following we will note the latent melting entropy $\Delta S_m$ [4].

The equilibrium condition (8) using (11) is splitted in two parts by defining the melting point $T_m^{'}$ of the strained film:

$$T_m^{'} = T_m - \frac{Em^2}{1-\nu} \frac{V^s}{\Delta S_m} \qquad (12)$$

and the reduced melting curve:

$$T_m^{'} - T = -\frac{N_{av}}{\Delta S_m} \frac{b^2}{\zeta} \left[ \Phi e^{-n_l/\zeta} - \Gamma e^{-n_s/\zeta} \right] \qquad (12')$$

Relation (12) implies since $\Delta S_m > 0$ *for melting that an epitaxial coherent strained film melts at a lower temperature than a strain-free film*. There is a shift of temperature which in the framework of elasticity theory is proportional to the misfit square $m^2$. This shift may be important: taking typical values for semi-conductors and metals ,$E=10^{11}$ ergcm$^{-3}$, $\nu=1/3$, $V^s=$

---

[4] A better approximation of (11) is when $\Delta C_m \neq 0$ but independent on T. Thus the following quadratic term [35] $-\Delta C_m(T-T_m) - \Delta C_m T\ln(T/T_m) \approx -\Delta C_m(T_m-T)^2/2T_m + O_3(\Delta T)$ adds to (11). Since $C_p^s < C_p^l < C_p^s$ for monoatomic elements one has at the high temperature limit $5R/2 < C_p^l < 3R$ so that a medium value of $\Delta C_m$ is -½ calmole$^{-1}$deg$^{-1}$. Neglecting this second order term brings for $\Delta T/T_m=0.1$ an error of 2% and for $\Delta T/T_m=0.5$ roughly 6%.

20 cm$^3$ mole$^{-1}$ the shift is 3.7, 15, 60 K for misfits of 1, 2 or 4 % respectively. For molecular deposits E=10$^{10}$ ergcm$^{-3}$ these values are ten times smaller.

At this point let us warn about some simplified treatments of the melting of stressed solids. For calculating (P,T) diagrams one uses the very valid procedure where the melting equilibrium s $\leftrightarrow$ 1 is given as a function of T and P the hydrostatic pressure surrounding both liquid and solid. One seeks for the solutions of the equation $\Delta G(T,P) = \Delta U_m(T) - T\Delta S_m(T) + P[V^l(T,P) - V^s(T,P)] = 0$ or with the approximation $\Delta c_m = 0$ (see footnote 4) $T_m^{'} = T_m - P[V^s(T,P) - V^l(T,P)]/\Delta S_m$ where $T_m^{'}$ is the melting point at pressure P, $T_m$ that one at P=0. Some authors [36-40] extend this approach to stressed solids by defining (i) a real hydrostatic pressure P=P$^l$ exerted on the liquid, (ii) a hypothetic "mean pressure" on the solid $\overline{P}^s = -\frac{1}{3}\delta_{ij}\sigma_{ij}$ where $\sigma_{ij}$ is the stress tensor, (iii) they write $\Delta S_m T_m^{'} = \Delta S_m T_m - \overline{P}^s V^s(T, \overline{P}^s) + P^l V^l(T, P^l)$. In the case the solid film s is epitaxially strained (appendix II) by $\varepsilon_{11} = \varepsilon_{22} = m$ so that appears in s an in-plane stress $\sigma_{11} = \sigma_{22} = Em/(1-\nu)$ and a normal stress $\sigma_{33} = -P^l$ one has when $P^l = 0$ and $\overline{P}^s = -2Em/3(1-\nu)$: $T_m^{'} = T_m + \frac{2}{3}\frac{E}{1-\nu}m\frac{V^s(T,0)}{\Delta S_m}\left[1 + 2m\left(\frac{1-3\nu}{1-\nu}\right)\right]$. From this result one sees that the second rhs term is composed of a leading term in m and a m$^2$ correcting term for Poisson effect. The same numerical data as above produce for a compressive misfit m=-1% a shift of roughly 240K of the melting point towards lower temperature instead of 3.7 K with (12). Misfits of –3% would bring layers (as Ge/Si(100)) to melt at room temperature or for systems with m>0 (tensile misfit) to become fire proof ! In [36-39] such values are seriously discussed for the system InAs/GaAs (m=-7%) and in [40] for the strain effect in surface melting of Ge/Si(100). In fact these predictions are wrong since the definition of the "mean pressure" in a solid has no physical meaning. More precisely such a "mean pressure" does not work during the melting process contrary to what is written in the above-mentioned point (iii).

Coming back to relation (12') at equilibrium, it means that some liquid layers can exist below or above $T_m^{'}$ according to the sign of the second member of (12') that implies either *premelting, overheating or both*.

**II.1.1/ Melting of semi-infinite solids (fig 1a): "Asymptotic premelting"**

This is the well known usual case performed without straining the solid so that m=0 and $n_s \to \infty$ so that $T_m' = T_m$ from (12). From (12') the number of equilibrium layers as a function of temperature (or melting curve (mc)) thus reads:

$$\frac{n_l^{eq}}{\zeta} \approx \ln\left[-\frac{N_{av}b^2}{\Delta S_m}\frac{\Phi}{T_m' - T}\right] \qquad \textbf{(13)}$$

Thus the equilibrium number of liquid layers increases asymptotically when $T \to T_m = T_m'$. This solution has only sense for $\Phi < 0$ that means (from (9)) *when the solid surface is wetted by its own melt* or, in terms of effective forces, that interface l/s is pushed away from the liquid surface. This is a stable equilibrium since from (8) $\left.\partial^2 G/\partial N_l^2\right|_T = -N_{av}b^2\Phi/\zeta$ is positive.

From (13) the surface remains dry $n_l^{eq}=0$ up to $T_s = T_m - \frac{N_{av}b^2}{\Delta S_m}\frac{|\Phi|}{\zeta}$ . In the case of Pb or Cu, $\Phi$=-20 erg cm$^{-2}$, b= 3.10$^{-8}$ cm [8]. When $\zeta$=1 there is $T_m$-$T_s$= 100 K and the successive liquid monolayers n install at 100e$^{-n}$ that means 37 K for one monolayer, 13 K for two layers and a very close approach to $T_m$ (4.10$^{-3}$ K) for 10 monolayers. Shifting $T_m$ to $T_m'$ by straining the solid s, see (12), may be a useful tool. Coherent epitaxy of s on a substrate S may be the most convenient practical solution but requires thin stable films where new proximity effects occur, which we approach in II.1.2.

**Variant: One side interfacial premelting**

For several experimental reasons it may be desirable to put on the surface of the thick solid s a cover glass (cg). Optical thickness measurements of premelted layers have been done by this means [9, 41,42]. Since the interfaces are different the surface free enthalpy (7) has to be changed. In appendix III we derive formula X which gives this new function of $n_l$ for a thick coverglass. Operating as in II1 and II4 instead of (13) one obtains the melting curve (mc):

$$n_l^{eq}/\zeta_l = \ln\left[-\frac{N_{avo}b^2}{\Delta S_m}\frac{\Phi - \left(\beta_{l/cg} - \beta_{s/cg}\right)}{T_m' - T}\right] \qquad \textbf{(13')}$$

that means the wetting factor $\Phi$ is changed by the adhesion properties of the coverglass.

     * In the case $\Phi < 0$ and when the liquid layer l adheres more on the coverglass that does the solid s, that means $\beta_{l/cg} > \beta_{s/cg}$ the number of equilibrium premelted layers

$n_l^{eq}$ according to (13') increases compared to the free liquid surface (13). In the glass covered surface melting studies of diphenyl [9] and ice [41] high number of liquid layers have been detected.

      * In the case of $\Phi>0$ where the solid is not wetted by its liquid, provided $\beta_{l/cg}-\beta_{s/cg}>\Phi$ premelting takes place and the stability criterion $\partial^2 G/\partial N_l^2\big|_T>0$ is satisfied. Here owing to the presence of the coverglass the liquid l has been "forced to wett" its own solid. Let us remark that such *forced wetting extends interestingly the realm* where surface melting can be studied by changing the nature of the cover glass. In [42] the coverglass has been grinded and its wettability has also been changed by surfactants.

This premelting is in fact an interfacial premelting in between two condensed phases but since the atoms of the coverglass do not participate directly in the melting process one should qualify it as one-sided. Grain boundary melting is an interfacial melting where both sides participate in the melting process.

## II.1.2/ Melting of a finite size solid (fig 2a):

The full expression (8) in which the term in $\Gamma$ becomes thus important the solid film is thin is now required to obtain the number of equilibrium layers. *By continuity we leave $\Phi<0$ but* consider the two cases $\Gamma>0$ and $\Gamma<0$.

$\Phi<0$ as said in II1.1 represents an effective repulsion force of the liquid surface upon the interface l/s. Now $\Gamma$ represents the interaction of l/s with the interface s/S where S is the epitaxial substrate. According to the definition (10) $\Gamma<0$ may be written $\gamma_l+\gamma_{ls}+\gamma_{sS}<\gamma_l+\gamma_{lS}$ what means that the l/s/S system is preferred to the l/S one. Therefore if $\Phi<0$ and $\Gamma<0$ the interface l/s is pushed away from both the liquid surface and the substrate S. These two forces thus may balance each other. In the case $\Phi<0$ and $\Gamma>0$ the two effective forces act in the same sense so that there is no balance and the l/s interface must be attracted to the substrate S. Clearly at each temperature T the third effective force of melting adds to the former ones. We will see that the case $\Phi<0$ and $\Gamma<0$ leads to a continuous increase of the number of liquid layers with temperature whereas in the case $\Phi<0$ and $\Gamma>0$ some instability from continuous to discontinuous behaviour occurs.

Consider the second derivative of G in respect to $N_l$ at constant T and N:

$$\partial^2 G/\partial N_l^2\big|_{T,N}=-\frac{N_{av}b^2}{\zeta^2}\left[\Phi e^{-n_l/\zeta}+\Gamma e^{-n/\zeta}e^{n_l/\zeta}\right] \qquad (14)$$

There follows the two new premelting cases we analyse more clearly in II1.2.1 and II.1.2.2.

## II.1.2.1. $\Phi<0$ and $\Gamma<0$: the surface induced continuous premelting and superheating:"astride melting"

From (14) there is $\partial^2 G/\partial N_l^2\big|_{T,N}>0$ for all values of $n_l$, so that in all the domain $0<n_l<n$, there is continuous stable melting from the dry point $T_s$ to $T_1$ the temperature where the last solid layer melted. From relation (12') there is with $n_l=0$, $n_s=n$:

$$T_s=T_m'+\frac{N_{av}}{\Delta S_m}\frac{b^2}{\zeta}\left[\Phi-\Gamma e^{-n/\zeta}\right]\approx T_m'-\frac{N_{av}}{\Delta S_m}\frac{b^2}{\zeta}|\Phi| \qquad \textbf{(15)}$$

the approximation being secured for a thick enough film for which thus $T_s$ does not depend upon $\Gamma$ but only upon the wetting $\Phi$.

At the same, from relation (12') with now $n_s=0$, $n_l=n$ there is:

$$T_l=T_m'+\frac{N_{av}}{\Delta S_m}\frac{b^2}{\zeta}\left[\Phi e^{-n/\zeta}-\Gamma\right]\approx T_m'+\frac{N_{av}}{\Delta S_m}\frac{b^2}{\zeta}|\Gamma| \qquad \textbf{(16)}$$

Thus for a thick enough film $T_1$ only depends upon $\Gamma$.

From (15) and (16) there is $T_s<T_m<T_1$. *The domain of continuous melting thus has to be called premelting when $T<T_m'$ or overheating when $T>T_m'$* (This situation remains for non strained layers, m=0, where then $T_m=T_m'$ according to (12))

This continuous melting at astride $T_m'$ is explicitly calculated by solving the quadratic equation obtained from (12') and (3):

$$\Phi e^{-2n_l/\zeta}+\frac{\Delta S_m\zeta}{N_{av}b^2}\left(T_m'-T\right)e^{-n_l/\zeta}-\Gamma e^{-n/\zeta}=0 \qquad \textbf{(17)}$$

At $T=T_m'$ from above there is

$$n_l\big|_{T=T_m'}=\frac{n}{2}-\frac{\zeta}{2}\ln\left(\frac{\Gamma}{\Phi}\right) \qquad \text{and} \qquad dn_l/dT\big|_{T=T_m'}=\frac{\Delta S_m\zeta^2}{N_{av}b^2}\frac{\exp(n/2\zeta)}{2\sqrt{\Phi\Gamma}} \qquad \textbf{(17')}$$

This inflexion point is close to half the number of total layers n and its positive slope increases exponentially with the total number of layers. In figure 3 this melting curve of continuous astride melting is schematically drawn.

## II.1.2.2. Φ<0 and Γ>0: the surface induced two stage premelting: "boosted premelting"

In this case, it can be seen from (14) that $\partial^2 G / \partial N_l^2 \big|_{T,N}$ may have positive or negative values and that $\partial^2 G / \partial N_l^2 \big|_{T,N} = 0$ for:

$$n_l^* = \frac{n}{2} - \frac{\zeta}{2} \ln\left(\frac{\Gamma}{|\Phi|}\right) \quad \text{where} \quad dn_l / dT \big|_{n_l^*} = \infty \qquad (18)$$

This singular value of $n_l$ can be smaller or greater than $n/2$ according to the value of $\Gamma / |\Phi|$. More precisely $n_l^* < n_l$ (resp. $n_l^* > n_l$) for $\Gamma / |\Phi| > e^{-n/\zeta}$ (resp. $\Gamma / |\Phi| < e^{-n/\zeta}$).

* For $0 < n_l < n_l^*$, equation (17) has two solutions. A stable one $n_l^{stable} < n_l^*$ where $\partial^2 G / \partial N_l^2 \big|_{T,N} > 0$ and an unstable one $n_l^{unst.} \geq n_l^*$ where there is $\partial^2 G / \partial N_l^2 \big|_{T,N} \leq 0$. In figure 3 (left to $T_m^{'}$) the two branches, the stable and the unstable one (dotted) are reported. They continuously meet close at $n_l^{stable} = n^{unst.} = n_l^*$ that means at the temperature

$$T^* = T_m^{'} - \frac{N_{av} b^2}{\Delta S_m \zeta} 2 \sqrt{|\Phi \Gamma|} e^{-n/2\zeta} \qquad (19)$$

obtained by injecting (18) in (12') and (3).

Since at this point $\partial^2 G / \partial N_l^2 \big|_{T,N} = 0$ any further increase of temperature produces an irreversible first order melting at $T^* < T_m^{'}$ as given in figure 3 by the heavy curve. Therefore when Φ<0,Γ>0 there is *a two stage premelting*. The first stage roughly concerns half the film which continuously melts. The second half corresponds to a first order melting at $T^* < T_m^{'}$, $T^*$ being given by (19).

* For $n_l^* > n$, the unstable solution $n_l^{unst.} \geq n_l^* > n$ obviously has no more meaning and there is only a stable one according to (17) corresponding thus to a continuous premelting from $T_s$ given by (15) to $T_l$ given by (16). Such a case could be encountered for systems with vanishing Γ. In this case the solution of the equation (12') is nothing else than (13). The number of equilibrium liquid layers thus is the same that for semi-infinite solids but obviously limited to $n_l^{eq} = n$. Thus for Γ=0 the curve $n_l^{eq}(T)$ fits the curve obtained for semi-infinite solids but is truncated at $n_l^{eq} = n$ (see the middle curve in figure 3a and the arrow at $n_l = n$.). As a

consequence, the $n_s$ solid layers become all liquid for $T_l = T_m^{'} - \dfrac{N_{av}}{\Delta S_m} \dfrac{b^2}{\zeta} |\Phi| e^{-n/\zeta} < T_m^{'}$ whereas for a semi-infinite solid all the solid layers only melt at $T = T_m^{'}$.

## III Discussion

### III.1. Premelting-overheating

The energetic interaction of the l/s interface located in between the interfaces v/l and s/S (see fig.4) is characterised by $\Phi$ and $\Gamma$. The effective forces on l/s read from (7)

$$f_{ls} = -\left( \dfrac{\partial G^{surf}}{\partial n_l} \bigg|_{n_s} \right) = -\left( \dfrac{\Phi}{\zeta_l} e^{-n_l/\zeta_l} - \dfrac{\Gamma}{\zeta_s} e^{-n_s/\zeta_s} \right).$$ In figure 4a where $\Phi < 0$ and $\Gamma < 0$ they oppose

each other, in figure 4b they work in synergy. Two melting regimes depicted in figure 3 result.

When the forces oppose ($\Phi < 0$ and $\Gamma < 0$) premelting starts at $T_s < T_m^{'}$ (15) but slows down when progressing in temperature (in respect with curve 1 in fig 3 where $n \rightarrow \infty$ drawn for comparison). Roughly half of the solid layers does not melt at $T \approx T_m^{'}$ (17'), several resist to melt up at T close to $T_l$. There is a tendency to retain solid layers s near to the substrate S. Reversing the pathway, starting from a liquid film at $T_l > T_m^{'}$ (what requires that the liquid l wets S), epitaxial solidification of s/S starts at $T_l$ and proceeds reversibly, passing $T_m^{'}$ and completes at $T = T_s$. In figure 5 we illustrate numerically for a ten layers system, how $\Phi$ and $\Gamma$ act on the melting curves all astride $T_m^{'}$.

Fig 5a depicts how for a given $\Phi < 0$ value the overheating zone is increased by the substrate repulsion $\Gamma$ the premelting zone being insensitive to $\Gamma$. At contrary in figure 5b we show how the premelting zone increases with the wetting $\Phi$ and the heating zone is insensitive to $\Gamma$.

When the forces on the interface l/s act in synergy ($\Phi < 0$, $\Gamma > 0$) (see figure 4b) increasing the temperature makes premelting to start continuously (left part of fig. 3) but speeds up ( with respect to curve 1 valid for $n_s \rightarrow \infty$ ) and when coming around the half melted solid due to the attraction ($\Gamma > 0$) of the substrate at $T^*$ just below $T_m^{'}$ (19) the first order melting takes place at constant $T^*$. Figure 6a shows how the premelting zone is insensitive to $\Gamma$ (only the unstable parts $n_l^{max}$, dashed curves, depend on $\Gamma$). Figure 6b shows the effect of wetting on the premelting ( insensitivity of $\Phi$ on the unstable solution $n_l^{max}$).

Reversing the pathway in the case $\Gamma > 0$ by starting with a liquid film at $T > T_l$ (fig 3) (what requires that the liquid l wetts the substrate S) attaining $T = T^*$ there is an activation barrier to overcome due to the discontinuities of $n_l$ at $T^*$, $n_l = n$ and $n_l = n_l^*$. It is $\Delta G_{l \to s}(T^*) = G(T^*, n_s = n - n_l^*, n_l = n_l^*) - G(T^*, n_s = 0, n_l = n)$ with $n_l^*$ given by (18). In appendix IV we show that $\Delta G_{l \to s}(T_l) \approx N_{av} b^2 \Gamma(n/2 + 1)$. So each successive solid layer has to overcome (in excess to the 2D nucleation barrier) the barrier $\Delta g_{l \to s}(T^*)/kT = \dfrac{3}{2} \Gamma b^2 / kT^*$ *due to the substrate S repulsion* since $\Gamma > 0$. Even for $\Gamma = 100$ ergcm$^{-2}$ and T=1000K this barrier has a high probability to be jumped.

### III.2. Which system exhibits which melting phenomena

We have to precise $\Phi$ and $\Gamma$ characteristics of existing systems.

#### III.2.1. $\Phi$ values: their determination

$\Phi$ values only concern the deposit and its self-wetting ability by its liquid secured when $\Phi < 0$. Pluis et al [11] in their paper have called $\Phi = -\Delta\gamma$ and have collected semi-quantitative data from Miedema et al.[43-45] upon $\gamma_s$, $\gamma_l$ and $\gamma_{sl}$ at the melting point for pure metals. They are empirical and average data ignoring anisotropy effects which in fact are small for metals (less than 4 %) what we know from the equilibrium shape of metals [46]. Miedema [43-45] collects first surface energies at $T_m$ [43] of roughly 30 clean liquid metals, the only clear measurable quantity and finds that they scale linearly with the vaporisation energy per unit molecular surface. For more general type of bonding this relation is called Stefan's rule (see K.Wolf [47]) amended by Skapsky [48]). By electronic empirical considerations, Miedema [44] finds that at 0 K there is the mean value relation $\left(\gamma_s V_s^{2/3}\right)_{0K} = 1.13 \left(\gamma_l V_l^{2/3}\right)_{0K}$. The $\gamma_l$ and $\gamma_s$ values are listed in Pluis [11]. For non metals we give in appendix V a rationalisation of the ratio $\left(\gamma_s / \gamma_l\right)_{T_m}$ we will use when the necessary data are needed.

The other quantity necessary for having insight how the liquid wets its solid is $\gamma_{sl}$ or ($\beta_{sl}$). For this purpose Miedema [45] follows Ewing'elegant procedure [49, 50] where $\gamma_{sl}$ is said to be the sum of an enthalpy term $\gamma_{sl}^l = k \dfrac{T_m \Delta S_m}{V^{2/3}}$ and a configurational entropy term $\gamma_{sl}^{II} = -\dfrac{T_m \Delta S_c}{V^{2/3}}$. The first enthalpic term is just a fraction k of the areal transition enthalpy across the sl interface, $k = (Z - z)/2Z$ being the fraction of liquid molecules a surface atom sees with Z the bulk coordination and z the in-plane coordination in the surface layer of the solid. For the second term $\Delta S_c < 0$ is the deficit of entropy of the liquid near the surface due to its layering ability. This second term thus describes how molecular disorder of the liquid differs when approaching the solid surface (supposed to be perfectly flat). Ewing [49,50] calculates this entropy deficit from the radial distribution function in the bulk liquid experimentally determined by X ray scattering. At a last resort this simplification is probably not too bad for the atoms considered in the compact crystal face since measurements on real surfaces have not really be done yet systematically. The entropy deficit listed in [45] table II lies inside $-0.8$ cl.mole$^{-1} < \Delta S_c < -0.3$ cl.mole$^{-1}$. The $\gamma_{sl}$ values with their $\gamma_{sl}^{I}$ and $\gamma_{sl}^{II}$ components are listed in table 3 and taken over by Pluis [11].

Among 33 elements where $\Phi$ data could be considered critically [43-45] 18 of them have negative $\Phi$ values with a peak at $\Phi = -25$ ergcm$^{-2}$, in the range $-50 < \Phi < -5$. The other 15 solid elements are predicted not to be wetted by their own melt, their positive values of $\Phi$ are largely spread inside $5 < \Phi < 180$ ergcm$^{-2}$. Table I gives in the second line the $\Phi < 0$ values of these elements. Premelting could be studied effectively on Al, Cu, Ga and Pb on several crystallographic orientations. On clean (0001) faces of Cd and Zn incomplete wetting is observed [51] with about a 30° contact angle. This may be the case for several other elements of table I when their most compact faces are considered (singular faces). In the third line of table I one sees that all $\widetilde{\Phi}$ values are positive. This is in fact true for all the 33 elements reported by [11]. In figures 1a and 1b where s and l have been exchanged $\widetilde{\Phi} < 0$ means no surface freezing. The author [11] could confirm from literature that none of theses liquid elements show surface induced freezing. Let us see why surface melting and surface freezing exclude mutually in the case of metals. Since by definition $\Phi = 2\gamma_l - \beta_{sl}$ and $\widetilde{\Phi} = 2\gamma_s - \beta_{sl}$ there is $\Phi = \widetilde{\Phi} - 0.26\gamma_l$ when we apply the mean numerical relation $\gamma_s = 1.13\gamma_l$. As a consequence the condition $\Phi < 0 < \widetilde{\Phi}$ or even $\Phi < \widetilde{\Phi} < 0$ necessarily is satisfied (fig. 1a, fig.2a). However $\widetilde{\Phi} < 0 < \Phi$ or $\widetilde{\Phi} < \Phi < 0$ for surface freezing (fig. 1b, fig 2b) are not necessarily satisfied. Miedema' relation $\gamma_s > \gamma_l$ in fact implies that the liquid surface and the solid surface are similarly relaxed what may be the case for globular molecules. This is not true for linear or sheet molecules which may have an higher orientational order at the surface than in the bulk so that the result is to increase $\gamma_l$ to $\gamma_l + \Delta\gamma_l$ and reverse the inequality. Known prefreezzing liquids [5,6] have such a surface organisation.

### III.2.2. Γ values: the lack of data

Γ values have to be known since necessary to predict what type of premelting occurs (for a given system with $\Phi<0$). In fact since $\widetilde{\Phi}$ and $\gamma_{ls}$ values are roughly known for these metals since Γ can be written $\Gamma = \widetilde{\Phi} + \beta_{lS} - \beta_{sS} = \gamma_{ls} + \gamma_{sS} - \gamma_{lS}$ the only missing data are $\beta_{lS} - \beta_{sS}$ or $\left(\gamma_{sS} - \gamma_{lS}\right)$ characterising the adhesion (or interfacial) energies of binary systems A/B. The subscripts l and s are valid for material A which is the deposit and S the subscript for material B. In the case $\widetilde{\Phi}>0$ ( or $\gamma_{sl}>0$) since $\beta_{lS} - \beta_{sS} <0$ (or $\gamma_{sS} - \gamma_{lS} <0$) which is the very general property that a liquid A adheres less on a substrate S than does the solid A especially when the interface sS is coherent. Therefore $\widetilde{\Phi}$ and $\beta_{lS} - \beta_{sS}$ compete each other so that Γ is either positive or negative. However the approximations to calculate such differences have to be handled with care.

Following Miedema' scheme [45] four successive approximations are made. **i)** interfacial energy $\gamma_{AB}$ is divided in three parts, the two "physical parts" defined above $\gamma^I_{AB}$, $\gamma^{II}_{AB}$ and the third one the "chemical part" $\gamma^{III}_{AB}$ making the very distinction of the two components of the binary system AB with $\gamma^{III}_{AA}=0$. Writing explicitly with evident notations:

$\Gamma = \left(\gamma^I + \gamma^{II}\right)_{ls \atop AA} + \left(\gamma^I + \gamma^{II} + \gamma^{III}\right)_{sS \atop AB} - \left(\gamma^I + \gamma^{II} + \gamma^{III}\right)_{lS \atop AB}$. Further simplifications are made:

**ii)** the chemical parts coming from the heat of solution of alloys, the distinction between sS and lS can hardly be made, the best is to write $\gamma^{III}_{sS \atop AB} = \gamma^{III}_{lS \atop AB}$. **iii)** the configurational entropy part of a liquid A meeting the solid surface A being taken from the bulk radial distribution of A the true nature of the surface A or B is of no matter and $\gamma^{II}_{sS \atop AA} = \gamma^{II}_{lS \atop AB}$. Furthermore $\gamma^{II}_{sS \atop AB}=0$ for a coherent interface. **iV)** the physical enthalpy parts making no distinction about A and B one writes $\gamma^{I}_{lS \atop AB} = \gamma^{I}_{lS \atop BB}$. Furthermore $\gamma^{I}_{sS \atop AB}=0$ per essence. Finally all these crude approximations lead to: $\Gamma = \gamma^{I}_{ls \atop AA} - \gamma^{I}_{lS \atop BB} = \dfrac{Z-z}{2Z}\left[\left(\dfrac{T_m \Delta S_m}{V^{2/3}}\right)_A - \left(\dfrac{T_m \Delta S_m}{V^{2/3}}\right)_B\right]$ so that one infers that mostly, $\Gamma<0$ since premelting of A can be done only on a substrate B when $T_m\big|_B > T_m\big|_A$. The weakest points of these approximations are i) tripartition , then points (ii) and (iV) so that some hope remains to meet systems with Γ>0.

### III.2.3. Γ determination: a proposal

May be the best solution is to trust *on direct experiments*, either on the premelting experiments where Φ and Γ data can be collected (see in this paper sections III4, VI3 and V3 as experimental examples) or on direct measurements of some ingredients of Γ. Supposing that from above $\widetilde{\Phi}$ (or γ_{ls}) is known their remains to know β_{lS}, β_{sS} (or γ_{lS}, γ_{sS}). To know the values of $2\gamma_s - \beta_{sS}$ and $2\gamma_l - \beta_{lS}$ would be as helpful as well as the necessary Φ value so that surface melting occurs when all these quantities are negative (see fig 2a). As a consequence neither contact angle measurements of the liquid l on the substrate S are of any help, contact angle being zero, nor measurements of 3D equilibrium shapes of crystals on S since the only stable states are the 2D wetting layers of s on S.

We propose that the determination of $2\gamma_s - \beta_{sS}$ <0 has to be done by measuring directly the thickness of the solid wetting layers. One puts in a closed isothermal empty volume the solid of surface S at the same horizontal level as the solid s. At T<$T_m^s$ where the vapour pressure of s is high enough $n_s$ molecules transfer on S as successive monolayers of s. When no mixing of s and S takes place the free enthalpy balance is written easily. Considering the coherent epitaxy on s/S, the essential limitation of the number of equilibrium layers is the strain energy surface density $Eam^2/(1-\nu)$ per layer (see formula (36) we apply in subsection VI3). When measuring the number of layers $n_s$ at equilibrium: $\left| 2\gamma_s - \beta_{sS} \right| = \left[ \dfrac{Em^2}{1-\nu} + 2\rho g \right] \xi_s a e^{n_s/\zeta_s}$ . When m→0, the potential energy difference $2\rho g n_s a$ in the gravity field of the final surfaces s and S of equal area becomes leading and very great thickness are expected. The parameter ζ_s may be approached by using the fact that misfit is the most sensible parameter versus temperature. For glissile epitaxies (see V) where strain is motivated by surface stress ($s_s+s_{sS}$) equilibrium strain ε_{eq} decreases as $n_s^{-1}$ according to (27). The above-mentioned relation is valid when to $m^2$ constant is substituted $\varepsilon_{eq}^2 \propto n_s^{-2}$. For a same $2\gamma_s - \beta_{sS}$ value a much greater number of equilibrium wetting layers are obtained. For a glissile system, this number is however smaller than for the case where gravity is the only limiting factor.

The determination of $2\gamma_l - \beta_{lS}$ <0 can be determined with a similar isothermal transfer system but at T>$T_m^s$. Clearly gravity is only the factor limiting the thickness (see [52]).

### III.3. van der Waals interacting interfaces

Up to now we considered the exponential decay of neighbour interactions of layers. Let us consider other interactions. Suppose all species interact according to $r^{-6}$ dispersion forces which owing to the fluctuating electromagnetic field are quite very general asymptotic forces valid for r>>a in condensed matter [52]. The surface excess free enthalpy for our planar system (fig 2a) given by V in appendix III reads instead of (7):

$$G^{surf}\left(n_s, n_l\right) = N_{avo}\left\{2\gamma_s + (2\gamma_l - \beta_{ls})\alpha\sum_1^{n_l} n^{-3} + (2\gamma_s - \beta_{sS})\alpha\sum_1^{n_s} n^{-3} + (\beta_{ls} - \beta_{lS})\alpha\left(\sum_1^{n_s} n^{-3} - \sum_1^{n} n^{-3}\right)\right\} \quad \textbf{(7')}$$

After some simplifications (as in II1) the first discrete derivative of (7') at constant n=$n_s$+$n_l$ brings to the melting curve similar to (12'):

$$\Delta S_m\left(T_m^{'} - T\right) + N_{avo}b^2\alpha\left[\Phi n_l^{-3} - \Gamma(n - n_l)^{-3}\right] = 0 \quad \textbf{(12'')}$$

with $n_l=n_l\big|_{eq}$ and $\Phi$,$\Gamma$ having the same meaning as in (9) and (10). Stability condition $\partial^2 G/\partial n_l^2\big|_n$ >0 thus requires (instead of (14)):

$$\left((n - n_l)/n_l\right)^4 < -\Gamma/\Phi \quad \textbf{(14')}$$

One distinguishes the same two stable premelting behaviour $\Phi$<0, $\Gamma\lessgtr0$ we called previously boosted premelting for $\Gamma$>0 (II.2.2) and astride melting for $\Gamma$<0 (II.2.1) both cases degenerating in the usual asymptotic premelting when the solid s becomes thick $n_s\to\infty$ .

The melting curves are similar to those schematically given in figure 3 and figures 5-6 for the exponential interaction with $\zeta\geq$1 excepted there is no more finite temperature $T_s\neq0$ . Below some temperature $T_s$ the surface should be dry $n_l\to0$ but the *asymptotic $n^{-3}$ law does not allow it*. This is quite unphysical and contradicts experiments and molecular simulations specially valid at low coverage [53]. A similar inaptitude happens at the other end of the melting curve (12'') where for the astride melting $\Gamma$<0 the last "solid atoms" do not transform in liquid ones. For these reasons we maintain the formulation (12) with exponentials where one may add asymptotic $r^{-3}$ tails if required.

Experiments where size effects could be approached through the force of circumstances are those of epitaxial adsorbed gases [18-23] that means precisely so-called Van der Waals systems. Studied still in the fifties [54-56] such systems show step adsorption isotherms characteristic of well defined surfaces and which are the sign of layer by layer growth. A discrete succession n of first order gas-solid 2D condensation at constant T takes place at well defined reduced chemical potentials of the vapour $kT\ln(P_n/P_\infty)<0$ that means at undersaturation $P<P_\infty$. Such a behaviour is well assessed for spherical or quasi-spherical molecules (noble gases, $CH_4$, $CF_4$ etc…) on lamellar crystals whose dominant surface is ideally flat and perfect as graphite, $MoS_2$, $CdI_2$ etc…. Solid epitaxial films of tunable thicknesses are then equilibrated with their vapour pressure. Approaching their melting point $T_m$ surface premelting could be observed [20].

### III.4/ An exemple: Coherent $CH_4$/MgO epitaxial system of astride melting:

A series of studies [19,21,22,56-58] on $CH_4$ ($CD_4$), melting point $T_m$=90.7 K (89.7 K) epitaxially grown on (001) MgO as fcc (001) $CH_4$ layers with $2[1\bar{1}0]$ MgO in parallel orientation with [100] $CH_4$ ($CD_4$). Up to five or more solid monolayers can grow. Beyond 3D crystals appear either due to inavoidable capillary condensation on the powder or due to some epitaxial strain. The natural misfit $m=(a_{MgO}\sqrt{2}-a_{CH_4,3D})/a_{CH_4,3D}$ varies in the temperature range 50-90 K from nearly zero to −1.5% essentially due to the thermal expansion of the deposit (that one of MgO being 30 times smaller). We take $a_{MgO}$=4.207 Å, $a_{CH4,3D}$=5.865(1+3.3 $10^{-4}$T) Å [22]. At roughly 50 K *a thick solid film would be therefore non strained when commensurate with MgO*. If temperature goes up to 90 K, if the deposit remains coherent, it becomes strained up to $\varepsilon_{//}$= -1.5 %. In fact the measured expansion of the film is constant within the error bars 0.003Å so that one has to conclude that the film is *at least coherent* with MgO(100) in all this temperature range. Its maximal compressive strain due to its hindered dilation varies linearly from $\varepsilon_{//}$=0 to $\varepsilon_{//}$=-1.5 % in the temperature range of 50-90 K. Taking $E/(1-\nu)\approx 1.9\,10^{10}$ erg.cm$^{-3}$ (with $\nu$=0.4)[5] the maximal strain energy density is 4.2 $10^6$ erg.cm$^{-3}$. With $V^s$=33 cm$^3$ mole$^{-1}$ at 90K , $\Delta S_m$=2.48 cl.mole$^{-1}$ [60] relation (12) exhibits a maximal melting point shift $T_m'-T_m$=−1.8 K. Thus data of $CH_4$ and $CD_4$ can be roughly reported on the same temperature plot as done by [21-22]. Now another point can be discriminated: the formation of 3D-$CH_4$ crystals [21] mostly at 50 K and less at higher temperatures up to 90K is not due to strain (much too small) which excludes a Stranski Krastanov transition (also called dewetting transition) but they are due to spurious capillary condensation in-between the grains of the MgO powder.

---

[5] For $CH_4$ no measurements are available but for isomorphous rare gases we know $C_{11}$ and $C_{12}$ [59] so that we considered the value $\nu$=0.4 and $\chi_s$=3(1-2$\nu$)/E and $\chi_l$ are volumetric compressibilities of s or l. There is at $T_m$ , $\chi_l/\chi_s\approx 3$ for Ar and for $CH_4$ there is $\chi_l$=1.6 $10^{-10}$ erg$^{-1}$cm$^3$ [60] so that if we take the same ratio $\chi_s$=0.5 $10^{-10}$ deducing E/(1-$\nu$).

The quantities $n_l$ and $n_s$ have been measured by neutron diffraction [22] as a function of temperature in the range 50<T<95 K. Solid layers *have been detected at 5 degrees above the melting point*: $n_s$=2.5 for a n=5.8 thick film. This is very convincingly confirmed by Quasi Elastic Neutron Scattering (QENS) where from a broad pedestal of the liquid signal emerges a narrow peak due to the solid layers [21]. The 2D liquid layers have a molecular mobility as large as the molecules of the bulk liquid at the same temperature. Authors [21,22] claim that the solid layers, probably closest to the substrate and most influenced by its field, therefore persist above the melting point $T_m$. *This is clearly what we call the astride melting with $\Gamma<0$ in III.2.1.* The above-mentioned authors used the term of "presolidification" or "prefreezing" which is at least incomplete since there is forgotten the associated premelting. It is confusing too since by prefreezing there is now defined in literature what we represented in fig 1b and 2b. Surprising for the same authors was also to find a good logarithmic law [21,22] instead of some expected power law for van der Waals systems

$$n_l = -2.14 \log\left(\frac{T_m - T}{T_m}\right) - 0.73 \text{ with } T_m\text{=90 K} , \text{ T<}T_m$$

The measured points of both neutron techniques are essentially present inside 0<$n_l$<4. Of course only those points for T<$T_m$ could be considered in this log type representation. The total number of layers in the film lies around 6<n<8 for the various experiments where unfortunately they could not be really kept constant by increasing T (molecules go over the vapour phase or in 3 D crystals or 3D liquids in capillars ). Nevertheless writing the previous relation in the exponential form

$$\exp(-n_l/0.93) = 2.44\,10^{-2}(T_m - T)$$

we compare with (12'), taking the small temperature shift correction $T_m^{'} = T_m - 1.8K$ :

$$\exp(-n_l/\zeta) = \frac{\Delta S_m \zeta}{N_{avo} b^2 |\Phi|}(T_m^{'} - T) + \frac{\Gamma}{|\Phi|}\exp(-(n - n_l)/\zeta) \qquad \textbf{(12''')}$$

and identifies $\zeta$=0.93. When $n_l$<<n is small enough the first rhs term is the leading term so that one identifies $\Phi$=-3.7±0.2 erg.cm$^{-3}$. We took $\Delta S_m$=2.48 cal.mole$^{-1}$.deg$^{-1}$, $b^2 = \left(6\sqrt{2}/2.10^{-8}\right)^2$ cm$^2$ the area occupied by one molecule and $T_m^{'}$=88.2 K one obtains the dry temperature $T_s \approx 47$ K which lies close to a point of measurement 0<$n_l$<0.2 at 46 K. Parameter $\Gamma$ has to be found among the measurements at $T>T_m^{'}$. Since above the melting point at T=95 K, $n_l$=3.3, $n_s$=2.5 [21] so that one obtains $\Gamma$=-10.6±1.0 erg cm$^{-2}$. With the so-obtained data $\zeta$, $\Phi$ and $\Gamma$ we plot in fig 7 the melting curve (full line) according the two exponential representation (12). For clarity we do not put for $T<T_m^{'}$ the experimental points [21,22] on figure 7 which obviously lie on the full curve. However for $T>T_m^{'}$ we put as bars the observed persisting solid layers $n_s$ up to the highest point of measurement. We infer following the full line that at 140 K there persists about one solid layer probably splitting then in 2D islands we estimate to disappear at $T_l=T_m^{'}\dfrac{\Gamma N_{avo}b^2}{\Delta S_m^{'}\zeta}\approx204K$. On figure 7 is also plotted with the same data the melting curve calculated with the asymptotic power law $n^{-3}$ but, as seen, gives a bad representation at low and high temperature. At low and at high temperature where there are respectively a few liquid and a few solid layers the $n^{-3}$ law is no more valid.

In section VI we come back to this example of coherent epitaxy to calculate the local strain $\epsilon_{zz}$ in the solid layers and compare it with that one determined experimentally by [22].

## IV/ Surface stress effect on coherent epitaxy

The expressions (1) and (7) of the Gibbs free energy have been established without any consideration of surface stress. In other words we have neglected the energy spent to deform surfaces and interfaces before accommodation of A (s+l) onto its lattice mismatched substrate S. Let us recall that the work of deformation of an isotropic planar surface of orientation n, at constant number of surface atoms, may be written $dW_{def}=As_nd\epsilon$ where $s_n$ is the surface stress (here a scalar) of the surface n of area A and $d\epsilon$ is the isotropic in-plane deformation[6]. In the following we omit index n since we have to do here only with one type of orientation. However having to do with various interfaces of same orientations n but different nature i,j we will substitute such labels. The surface stress effect can thus be easily taken into account by adding to the Gibbs free energy of the system (1) the work of deformation of the solid surface and interfaces during accommodation of the solid + liquid film onto the substrate S.

At this point of the study it is very important to stress on the reference state used for the definition of the surface and interface quantities (Refer to appendix VI). In the Gibbs free energy (1,5-7) the reference is written in Lagrangian coordinates, that means the reference state is the non-deformed one. Therefore (see appendix VI) according to Shuttelworth' equation in Lagrangian coordinates there is $s_i=\partial\gamma_i/\partial\epsilon$ [63]. The supplement of surface energy

---

[6] For s >0 the surface layers tend to contract themselves and s is said to be tensile. For s<0 the surface layers tend to expand themselves and s is said to be compressive. For a recent tutorial paper on surface stress see [61] and for a comprehensive review [62].

is $2m\dfrac{\partial G^{surf}}{\partial \varepsilon}\bigg|_{\varepsilon=m}$ where $G^{surf} = N_{avo}\,g^{surf}$ is given by (7) and m=$\varepsilon_{xx}$=$\varepsilon_{yy}$ the in-plane

deformation. The adhesion energies $\beta_{ij}$ within(7) have to be converted by Dupré' relation (Appendix III, formula III) so that (7) reads (with $\zeta=\zeta_l=\zeta_s$)

$$g^{surf} = (\gamma_l - \gamma_s + \gamma_{sl})\left(1 - e^{-n_l/\zeta}\right) + (\gamma_s - \gamma_S + \gamma_{sS})\left(1 - e^{-n_s/\zeta}\right) + (\gamma_s - \gamma_S + \gamma_{lS} - \gamma_{sl})e^{-n_s/\zeta}\left(1 - e^{-n_l/\zeta}\right) + 2\gamma_S$$

Applying $s_i = \partial\gamma_i/\partial\varepsilon$ (i, s, l, S, $s_S$, $s_{lS}$) but noticing that $\partial\gamma_l/\partial\varepsilon$=0 since a liquid surface cannot be deformed at constant number of atoms, and noticing that $\partial\gamma_S/\partial\varepsilon$=0 since the very thick substrate S does not work during accommodation, there is:

$$\frac{\partial g^{surf}}{\partial\varepsilon} = \left[\Phi'\left(1 - e^{-n_l/\zeta}\right) + \Gamma_1'\left(1 - e^{-n_s/\zeta}\right) + \Gamma_2'\left(1 - e^{-n/\zeta}\right)\right] \tag{20}$$

with $\Phi' = s_{sl} - s_s$, $\Gamma_1' = s_{sS} + s_{ls} - s_{lS}$ and $\Gamma_2' = s_s - s_{ls} + s_{lS}$ (20')

Then surface and interfacial stresses only modify the factors $\Phi$ and $\Gamma$ in formulae (8, 9, 10) and the subsequent being concerned that now reads:

$$\Phi_I = \Phi + 2m\Phi' \text{ and } \Gamma_1 = \Gamma + 2m\Gamma_1' \tag{21}$$

A short discussion can be done on the qualitative effect of surface stress. First let us recall that constant misfit, whatever its sign, increases the elastic energy of the film and shifts, as discussed in II.1 the melting curves to lower temperature without altering their shape. Surface stress as shown by (21) acts on the shape of the $n_I^{eq}(T)$ curve by means of $\Phi_I$ and $\Gamma_I$ and depends upon the sign of the misfit m. From II123 one knows that $\Phi_I<0$ determines the premelting zone (fig 5b and 6b) and that $\Gamma_I$ when negative determines the overheating zone (fig 5a). From (21) it can be seen how misfit acts as a correcting term. In general surface stress $s_i$ is a positive quantity of the order of surface free energy $\gamma_i$ ( in absence of foreign adsorption) [62,64]. Interfacial free energies of solid/liquid roughly are five times smaller than the corresponding surface energies. Interfacial s/l stress is supposed to behave similarly. As a result of these estimates in (21) the corrective factors read $\Phi'$=-$s_s$+O($s_{ls}$) and $\Gamma'$=$s_{sS}$+O($s_{ls}$) so that from (21) there is:

$$\Phi_I \approx \Phi - 2ms_s \text{ and } \Gamma_I \approx \Gamma + 2ms_{sS} \tag{22}$$

Positive misfit thus will increase the premelting zone and eventually if $\Phi>0$ and small it will render wetting possible $\Phi_I<0$. Negative misfit at the contrary decreases the premelting zone rendering eventually $\Phi_I>0$ and thus annihilating premelting. Notice that for $\Phi$=-20 ergcm$^{-2}$, $s_s$=10$^3$dyncm$^{-1}$ (Cu, Pb) [64] $\Phi_I\approx$-40 or $\Phi_I\approx$0 ergcm$^{-2}$ for respectively m=1% and m=-1%.

Interfacial stress has an opposite but smaller effect ($s_{sS} < s_s$) on $\Gamma_I$ and thus on the overheating zone when $\Gamma < 0$ and $\Phi < 0$. One should come back to fig 5b, 6b where the melting curves are drawn for decreasing $\Phi$ up to $\Phi = 0$ and $\Gamma = cte$. $\Phi \approx 0$ typically is a case where no premelting but only overheating prevails when $\Gamma < 0$. When there is $\Gamma > 0$ nothing special happens at the surface around $T'_m$.

## V/ Non coherent epitaxies: glissile epitaxies

The coherent epitaxial films we treated in II-IV when acquiring a great enough thickness may release their elastic energy. Many type of defects may produce such a relaxation. The most studied defects are misfit dislocations. It is well known that above some critical thickness $n_s^c$, roughly depending upon the inverse of the misfit m and some stiffness ratio, dislocations suddenly introduce leaving the semi-coherent film with a residual misfit m'<m. The melting curves are thus shifted to a higher melting point $T'_m < T'_{m'}$. If dislocations are not too much hindered kinetically close to melting point their entrance is continuous, the residual misfit drops as $n_s^{-1}$ and the melting point $T'_m$ comes back to $T_m$ as $n_s^{-2}$.

*Non coherent epitaxies* act in a very different way. An extreme situation is the *perfectly glissile epitaxy* where no elastic accomodation in between the deposited film and its substrate is supposed to take place. The epitaxial films have not their natural crystallographic parameter and therefore are strained. When surface and interfacial stresses are properly taken into account the in-plane crystallographic parameter of the film is thickness dependent as well as the so-generated deformation ε and stress σ. Such kind of homogeneous models have been introduced for discussing wetting-non wetting behaviour [65], mechanical properties of thin films [66] and asymptotic stress of thin films [67]. They differ from the inhomogeneous models [68,69,23] ignoring surface stress but considering that the substrate strains inhomogeneously the solid film. We consider latter one consequently in section VI.

For our system (fig 2a) where the natural misfit m does not determine the strain, we introduce a virtual in-plane deformation ε of the glissile film, the Gibbs energy reads:

$$G(T, \varepsilon) = N_s \left[ G_s(0) + V^s \frac{E}{1-\nu} \varepsilon^2 \right] + N_l G_l(O) + N_{av} g^{surf}\left(\gamma_l(\varepsilon)\right) \qquad \textbf{(23)}$$

The equilibrium strain thus is obtained by minimising (23) in respect with ε. Thus there is

$$\varepsilon_{eq} = -\frac{1-\nu}{2Ea}\frac{1}{n_s}\frac{\partial g^{surf}(n_s,n_l)}{\partial \varepsilon} \qquad (24)$$

where we took $n_s = N_s a^2$ since the film is non coherent. The partial derivative has been calculated in (20) as a combination of surface stresses. The slab therefore is homogeneously in-plane strained (24) and $\varepsilon_\perp = -\frac{2\nu}{1-\nu}\varepsilon_{eq}$ but varying with size $n_s$ and $n_l$.

For calculating the equilibrium number of liquid layer $n_l$, one proceeds as in II1, injecting (24) in (23) and derivating in respect to $N_l$ at constant number N of atoms. When using inside $g^{surf}(\gamma_i(\varepsilon_{eq}))$ the developement of surface energies with respect to strain under the form:

$$\gamma_i(\varepsilon_{eq}) = \gamma_i(0) + \int_0^{\varepsilon_{eq}}\int_0^{\varepsilon_{eq}}\left(\frac{\partial \gamma_i}{\partial \varepsilon_{xx}}d\varepsilon_{xx} + \frac{\partial \gamma_i}{\partial \varepsilon_{yy}}d\varepsilon_{yy}\right) = \gamma_i(0) + 2\varepsilon_{eq}s_i \qquad (25)$$

From (23) calculating $\partial G(T,\varepsilon_{eq})/\partial N_l = 0$ one obtains similarly to (8), using (24) and (20) for the derivative $\partial \varepsilon/\partial n_l$ and $N_{av}b^2 a = V^s$, the melting curve:

$$\Delta S_m(T_m - T) - 3V^s\frac{E}{1-\nu}\varepsilon_{eq}^2 + N_{av}\frac{b^2}{\zeta}\left[(\Phi + 3\varepsilon_{eq}\Phi')e^{-n_l/\zeta} - (\Gamma + 3\varepsilon_{eq}\Gamma')e^{-n_s/\zeta}\right] = 0 \qquad (26)$$

where $\varepsilon_{eq}$ is given by (24), (20) and reads

$$\varepsilon = -\frac{1-\nu}{2Ea}\frac{1}{n_s}\left[\Phi'\left(1-e^{-n_l/\zeta}\right) + \Gamma_1'\left(1-e^{-n_s/\zeta}\right) + \Gamma_2'\left(1-e^{-n/\zeta}\right)\right] \qquad (26')$$

The definition of $\Phi$ and $\Gamma$ are given in (9) and (10) in term of surface energies at zero strain and $\Phi'$, $\Gamma_1'$ and $\Gamma_2'$ in (20') as a combination of surface stresses. Latter quantities are strain independent in contrast to volumetric stresses for which Hooke's law holds [70][7]. We used this property by writing (25).

Since to equation (26) is associated (26') the melting curves are distorted by the strain $\varepsilon_{eq}(n_l,n_s)$ in very a complex manner. For coherent epitaxies we have seen in (12) there is only a melting point shift towards lower temperatures and surface stress (see IV) shifted only the $\Phi$, $\Gamma$ characteristics.

**V.1/ Strain behaviour during melting:**

Let us see this more closely, looking first at the $\varepsilon(n_s,n)$ behaviour with (24) and (20). In figure 8 we draw $\varepsilon(n_s/n)$ for different thicknesses $n=10\ldots100$ taking $s_{ls}=s_{lS}=s_{sS}=200$ and typically $s_s = 10^3$ dyncm$^{-1}$ scaling with $E = 10^{11}$ ergcm$^{-3}$ according to the rule $3s/Ea=1$. $\varepsilon(n_s/n)$ is compressive when surface stress is positive. It is greater when the surface is dry ($n_s/n=1$) and also when there remains only a few residual solid layers. In between, the compressive strain has a minimum.

---

[7] When non linear elasticity has to be used then surface stress becomes strain dependent.

Let us notice that the curves of figure 8 have been interrupted at the left for $n_s<2,3$ since there surface stress defined as a macroscopic quantity looses its physical meaning. Nevertheless a film of n=10 layers has a strain -1.5%<ε<-0.7% a ten times thicker ones n=100, -0.15%<ε<-0.05% .

Interesting aspects around $n_s=n$ are revealed. For the dry film one has

$$\varepsilon_{eq}\left(n_s=n\right)=-\frac{1-\nu}{2Ea}\frac{1}{n_s}\left(s_s+s_{sS}\right) \qquad (27)$$

By thinning mechanically this film its compressive in-plane strain would increase hyperbolically (see fig. 7 dashed for n=10 on the rhs). However when becoming slightly wet its compressive strain decreases strongly. From (24) and (20) there is

$$\frac{d\varepsilon}{dn_s}\bigg|_{n_s=n}=\frac{1-\nu}{2Ea}\left[\frac{s_s+s_{sS}}{n^2}-\frac{s_s-s_{sl}}{n}\right]\approx-\frac{1-\nu}{2Ea}\frac{s_s-s_{sl}}{n}$$ to compare with the opposite and much

smaller slope on the hyperbola, $\dfrac{d\varepsilon}{dn}=\dfrac{1-\nu}{2Ea}\dfrac{s_s+s_{sS}}{n^2}$, $\left(s_{ij}\approx\dfrac{2}{10}s_i\right)$ of the dry film. Therefore when this surface becomes just wet the slope changes its sign and its absolute value is ten times greater when n=10. Clearly the continuous film model cannot account for the local change at $n_l=n$. The build up of the first liquid submonolayer behaves probably like an adsorption-condensation process. It is known [62, 64] that compressive stress (Δs>0) is often induced by the charge transfer from the surface to the adsorbed atoms. That means adsorbed electronegative species decrease surface stress and we associate them with the adatoms in the liquid state. The effect is strong since according to [62,64] surface stress $s_s$ often drops to zero at coverages as small as θ≈1/10 and becomes several times its but negative value at θ≈1/2. Its simplest form describing this behaviour may be tentatively written $s_s(\theta)=s_s\left(1-8\theta_l-16\theta_l^2\right)$. One sees that locally $\varepsilon\left(n-\theta_l\right)$ sweeps round at point $n_l=n$ when inserted in (27). Therefore also the appearance of the first melted layers is not properly accounted by theses macroscopic descriptions.

**V.2/ Melting curves**

Finally calculating melting curves with (26) for a n=15 layers thick film with the same standard data as in figure 8 corresponding in (26) to Φ'=-800, Γ'=200 dyn.cm$^{-1}$ and surface energy characteristics Φ=-30, Γ=+/-50 erg.cm$^{-2}$ one obtains respectively the continuous curves of fig. 9a 9b. There we plot also dashed curves representing some reference for non strained epitaxies ($dg^{surf}/d\varepsilon=0$ which brings back to relation (8) where m=0 instead of (26)). The main effect that is observed is that surface stress induced strain leads to important distortions of the melting curves. It reduces the premelting zones in both types $\Gamma>0$ or $\Gamma<0$ and reduces the overheating zone in the case $\Gamma<0$. Therefore if the distinguishing characters of both type of surface induced melting are preserved so that one can conclude by saying that in glissile epitaxies surface stress represses surface melting. This is however not true, *both repression or promotion* may occur according to the sign of surface stress. More generally, the elastic energy in (26) every time positive whatever surface stress promotes surface melting (as seen in coherent epitaxy in II1). The term $\varepsilon\Phi'$ in (26) whose sign depends on the sign of $(s_s+s_S)\Phi'$ when negative overwhelms the elastic energy and the wetting energy $\Phi$ so that it represses premelting as in the chosen example of figure 8. If however $(s_s+s_S)\Phi'$ is positive, the opposite happens and surface stress promotes premelting of glissile epitaxial films.

When looking closer around the melting point $T_m$ there are other qualitative changes quite different in both cases $\Gamma=\pm50$. In figure 9b where $\Gamma=+50$ a continuous premelting is relayed by first order premelting. Without strain the starting is at $n_l^*(l)=7.5-0.5$ and $T^*(0)<T_m$ given by (18) and (19) with the nominal values of $\Phi$, $\Gamma$ and $\zeta=1$. With strain the first order transition starts earlier at n*=5.3, $T^*<T^*(0)$ being given by the single zero of the second derivative of G, $\partial^2 G/\partial n_l^2\big|_{n^*}=0$. In figure 9a where $\Gamma=-50$ things suffer qualitative changes. The reference curve (dashes) shows the continuous premelting-overheating behaviour. The strain induced by the surface stress induces a first order melting starting at $n_l^*=5$, $T^*<T_m$ where obviously the second derivative $\partial^2 G/\partial n_l^2\big|_{n^*}=0$. However this second derivative G'' (see insert figure 9a) also vanishes for $n_l^{**}$ where ends this first order premelting which then is again relayed by a continuous overheating. At $T^*$ in between $n_l^*$ and $n_l^{**}$ the free enthalpy is flat but at $T<T^*$ exhibits a maximum between both minima at $n_l^*$ and $n_l^{**}$. Therefore in figure 9a at $T^*$ between $n_l^*$ and $n_l^{**}$ we substitute dots to the equilibrium curve. Clearly this peculiar behaviour is due to the complex changes (see figure 7) in the solid part of the film acting on $\Phi$, $\Gamma$ and their derivatives with respect to strain $\varepsilon$ (see (26)), (24) and (20)). For more negative values of $\Phi$ than in figure 9 the behaviour close to $T_m$ becomes qualitatively that of the reference curve.

**V.3/ Krypton/Graphite: a glissile epitaxy.**

At low temperature Kr grows from the vapour layer by layer on the basal plane (0001) of graphite [71]. In the range T=16 K to $T_t$=114.75 K the epitaxial orientation is (111)Kr//(0001) gr with $[1\bar{1}0]$ $a_{Kr}$ $\sqrt{2}/2//[21.0]a_{gr}\sqrt{3}$. No Stranski-Krastanov transition is observed by RHEED up to tens of monolayers [71] that means layer by layer growth is not relayed by three dimensional growth. This is the sign that the film is not severely strained. In the monolayer range around 55 K however the layer suffers a solid-solid transition of second order [72] interpreted as a 2D rotational static distortion (see [73] for such 2D diagrams). However when the second layer builds up the underlying Kr atoms move out from the graphite bonding sites and the film becomes quasi-autonomous, having its own crystallographic in-plane parameters attested by a distinct diffraction pattern with respect to the graphite. This is what is called an incoherent Kr-graphite interface (not a semi-coherent interface with dislocations). Nevertheless it is an epitaxial film since as above mentioned there is a strict azimuthal orientation. When comparing the in-plane parameters of the 3D juxtaposed phases in the epitaxial orientation, a mismatch $m=\left(a_{gr}\sqrt{3}-a_{Kr}\sqrt{2}/2\right)\!/a_{Kr}\sqrt{2}/2$ can be defined. Because of the thermal dilatation of Kr in this temperature range (the graphite parameter $a_{gr}\sqrt{3}$=4.256 Å remains quasi constant in this temperature range but $a_{Kr}^{16K}\sqrt{2}/2=3.991$ Å, $a_{Kr}^{115K}\sqrt{2}/2=4.124$ Å) this mismatch m is temperature dependant. Let us note that m is a measure of the in-plane strain only for coherent epitaxies and not for incoherent ones. Here the film glides rigidly over the substrate by a temperature change so that one call it glissile epitaxy. The denomination floating phase is sometimes used but in this context is misleading since the epitaxial orientation is strictly preserved during the gliding. The solid film is quasi-autonomous, so that its temperature and thickness-dependent strain has to be determined by precise X ray parameter determinations what has not yet be done up to now. Indirect strain evaluation however has been done. Dash and coworkers [20,23] observed heat capacity anomalies just beneath the melting point of Kr grown as multilayers on graphite. The heat capacity curves (hcc) plotted versus the reduced temperature $t=(T-T_m)/T_m$ (after subtraction of the blank and normal heat capacities) scales with derivatives of the melting curves (mc), $C(t)=K\dfrac{dn_l}{dt}\bigg|_n$ at constant number of layers $n=n_s+n_l$ so that to the maximum of a hcc corresponds the inflexion point of the mc. Experimentally systematic shifts of the different pre-melting curves towards lower temperature are observed (fig. 11a). The peaks behave as a nest of dolls, each peak shifting according the total number of atoms n. The authors [20,23] identified this behaviour as "strain assisted premelting". Strain energy calculations were modelled by a substrate z-attraction. In section VI we also analyse such inhomogeneous normal-strain but show that it is only a supplementary effect. Here we re-evaluate the authors [20,23] experimental results as due to homogeneous in-plane strain motivated by the four intrinsic interfacial stresses $s_s, s_{sl}, s_{sS}, s_{lS}$. With the $\Phi$ and $\Gamma$ parameters this is too much to come to a clear result so that we proceed in a heuristic manner. In former section V2 we had a general discussion about mc of glissile epitaxies (fig 9a,9b). In figure 9a where $\Phi, \Gamma < 0$ astride melting happens with a S-shaped mc so that a more or less symmetric hcc results. From figure 1 of [23] the measured hcc for Kr/graphite for respectively a total number of layers n=7.2, 8.7, 10.3, 11.8 show (fig 11a) that the peaks **(i)** are fairly symmetric so that we infer $\Phi = \Gamma < 0$, **(ii)** are not distorted so that $\Phi' = \Gamma' = 0$, **(iii)** can be brought back to origin. Owing to these simplifications (26) (26') can be rewritten in reduced variables $\Delta t_n(x) = t(x) - t_{el}(0)$ where x=n/2-$n_l$:

$$\Delta t_n(x) = -A_0 e^{-n/2\zeta} sh(x/\zeta) \qquad \textbf{(26'')}$$

with $A_0 = \dfrac{2N_{avo}b^2|\Phi|}{\Delta S_m T_m \zeta}$ and $t_{el}(0) = \dfrac{3N_{avo}}{\Delta S_m T_m} \dfrac{1-\nu}{E} \left(\Gamma_2'\right)^2 \left(\dfrac{1-e^{-n/\zeta}}{n}\right)^2$ $\qquad \textbf{(26''')}$

the shift depending upon n. $\Gamma_2' = s_{sS} + s_{sl}$ is the interfacial stress (21') which creates the elastic energy, the other quantities $s_s$ and $s_{lS}$ don't act independently since the imposed condition $\Phi' = \Gamma' = 0$ leads to $s_{lS} = s_{sS} + s_{sl}$ and $s_s = s_{sl}$. The sh term in (26'') gives a S-shape mc leading to a symmetric hcc. Now one can calculate the hcc as a function of the reduced variable t. Inverting (26'') gives:

$$x = -\zeta sh^{-1}\left[\dfrac{e^{n/2\zeta}}{A_0}\Delta t(x)\right] \qquad \textbf{(27)}$$

so that the hcc becomes:

$$C_n(\Delta t) = K\dfrac{dn_l}{dt}\bigg|_n = \left(\dfrac{K\zeta}{A_0}e^{n/2\zeta}\right)\bigg/\sqrt{1+e^{n/\zeta}\left(\dfrac{\Delta t}{A_0}\right)^2} \qquad \textbf{(28)}$$

where $A_0$ is given by (26'''). This is a symmetric peak centered around $\Delta t=0$ given by (26'''). The peak values $C_n(0) = \dfrac{K\zeta}{A_0}e^{n/2\zeta}$ taken from fig 1 [23] reported in figure 11a when semi-log plotted versus n gives a fair straight line (see figure 10a and stars) whose slope brings $\zeta=2.75$ and an intercept $C_{n=0}(0) = \dfrac{K\zeta}{A_0}$ =7.07 JK$^{-1}$. The peak width $2\Delta t_{1/2}(n)$ at half peak height is $\Delta t_{1/2} = \sqrt{3} A_0 e^{-n/2\zeta}$. From Dash'results one gains the number ratio A$_o$= 7.8 10$^{-3}$ and with (26''') an estimate of $\Phi = \Gamma$=-(0.17±0.06) erg cm$^{-2}$. We took the values b=4.124 Å, T$_m$=116 K, $\Delta S_m$=3.37 clmole$^{-1}$ [60]. The scaling factor is K=2 10$^{-2}$ JK$^{-1}$ but unfortunately cannot lead to an independent estimation of $\Phi$ since the unit area is not specified in the experimental work [23].

One has to evaluate also the peak shift t$_{el}$(0) according to (26'''). Figure 10b shows the appropriate fit giving a straight line passing by the origin. Using the value $E/(1-\nu)$=4.4 10$^{10}$ erg cm$^{-3}$ for a (111) plane (appendix II)from the elastic constants at 116 K [59] one earns the interfacial stress $|s_{sS}+s_{sl}|$=80± 10 dyncm$^{-1}$.

Finally in figure 11b our calculated hcc' facing the experimental one (fig 11a) shows a nice resemblance.

Let us collect and discuss shortly the physical quantities we deduced from our analysis. The number $\zeta$=2.75 refers to the (111) stacking so to a correlation length of some 9Å lying in between third and fourth nearest neighbour distance in Kr. The wetting factor $\Phi$ which is very small leads (10) to the adhesion energy $\beta_{sl}$=32.6±1 erg cm$^{-2}$ since at T=117 K there is $\gamma_l$=16.40±0.02 erg cm$^{-2}$ [60]. In appendix V we evaluate the ratio from an isotropic model at the melting point $(\gamma_s/\gamma_l)_{T=116\ K}$=1.30 so that $\gamma_s$=21.3 erg cm$^{-2}$ . There are calculations of two authors agreeing within ± 1% for noble gas crystals [64] where good potentials were available. For Kr, $\gamma_{(111)}$ =52.8 erg cm$^{-2}$, for (100) and (110) faces the values are respectively 4% and 8% greater so that one has a measure of the anisotropy . These values are valid at O K so that at higher temperatures several surface contributions reduce it. Zero point energy (5%), vibrational entropy (25% around T$_m$) and mostly surface roughness may reduce it all together by 40% (see for such temperature effects [46]).

Since $\Phi$=$\Gamma$ one has from (10) $\beta_{sS}$-$\beta_{lS}$=9.8 erg cm$^{-2}$ but we have not access to the separate adhesion energies. That solid Kr adheres better on graphite than liquid Kr does is quite expected from configurational entropy higher in the lS interface than in the sS interface even when latter is incoherent (see for more discussion section VI2). Finally the sum of interfacial stresses we deduced $|s_{sS}+s_{sl}|$=80±10dyncm$^{-1}$ is probably close to $|s_{sS}|$ since s$_{lS}$=s$_s$ and if small. The argument is that calculations on Kr(100) at O K [64,74] give s$_{(100)}$=-6 dyn cm$^{-1}$ that means a small value which may be similar for s$_{(111)}$.

Lastly we derive the in-plane strain as induced by the surface stress of the Kr layers. With (26') there is with the above Kr data at half melting $n_s=n/2$ with (20')

$$\varepsilon_{1/2}(n) = \pm 4.7 \, 10^{-2} \left( \frac{1 - e^{-n/\zeta}}{n} \right)$$ that means a quasi hyperbolic n-dependence as soon as n>2.

For the four increasing layer numbers n= 7.2; 8.7; 10.3 and 11.8 there is respectively $\varepsilon_{1/2}$=6.0 $10^{-3}$; 5.2 $10^{-3}$; 4.4$10^{-3}$ and 3.9 $10^{-3}$. When the solid becomes dry, these strains have to be divided by two. Clearly such glissile systems neither undergo a Stranski-Krastanov transition as quoted earlier in V3 nor have the tendency to dislocation-introduction during the growth since the strain induced by surface stress decreases with the thickness of the solid film.

## VI/ z-Inhomogeneity due to the finiteness of the film

In the foregoing sections we have seen that surface melting of finite size solids is governed by the wetting factors $\Phi$ and $\Gamma$ defining effective non-zero thermodynamic forces *between the various interfaces* when thermodynamical equilibrium is not reached ((8) and fig 4). Due to theses excess energies of the interfaces *also mechanical body forces act on each elementary slice of the slabs l and s*. Up to now we have neglected these forces especially in appendices I and II where we wrote the mechanical equilibrium conditions ignoring them. In a Gibbs treatment these forces have to be accounted for. Similarly to gravitational forces, these forces modify the mechanical equilibrium and create inhomogeneous stress and strain in the slabs. We will see that though very localised at the interfaces s/S and l/s, this strain inhomogeneity in the solid is measurable. However we will see that the premelting curves we discussed before are not sensibly modified by this strain inhomogeneity.

In order to justify the wetting-unwetting transition of noble gases film, inhomogeneous model slabs have still been treated by different approaches [68,69,23] and the so stored elastic energy calculated.

In the following we will define first the interfacial field, then the body forces which lead to inhomogeneous stress and strain, then calculate the elastic energy components with a clear distinction between coherent and glissile epitaxies. For this purpose we proceed in two steps.

* At first we calculate the excess field $\Omega_i(z)$ felt by a solid (i=s) or a liquid (i=l) layer located at a distance z above the substrate S in respect to the field felt by the same solid (or

liquid) layer in a continuous solid (or liquid) material. The calculations, developed in appendix VI, give for $n_S \rightarrow \infty$ (see formulae IV and VII for the slab of fig 2a):

$$
\begin{cases}
\Omega_l(z) = (2\gamma_l - \beta_{l/s})e^{-(z-n_s)/\zeta} + (\beta_{l/s} - \beta_{l/S})e^{-z/\zeta} & ; \quad for \quad n_s < z < n_s + n_l \\
\Omega_s(z) = (2\gamma_s - \beta_{s/S})e^{-z/\zeta} - (2\gamma_s - \beta_{s/l})e^{-(n_s-z)/\zeta}\left(1 - e^{-n_l/\zeta}\right) & ; \quad for \quad 0 < z < n_s \\
\Omega_S(z) = (\beta_{l/S} - \beta_{s/S})e^{-(|z|+n_l)/\zeta} + (2\gamma_s - \beta_{l/S})e^{-(|z|+n_s+n_l)/\zeta} - (2\gamma_s - \beta_{s/S})e^{-|z|/\zeta} & for \quad -\infty < z < 0
\end{cases} \quad \textbf{(28)}
$$

We see that the ingredients of these fields are surface and adhesion energies of l/s/S similar to those appearing in excess surface energy (7).

- In a second step we calculate the stress field induced by this excess field. For this purpose we assume that the previous excess field induces a body force field given by $af^i(z)=d\Omega_i(z)/dz$. The so induced stress is thus obtained by integration of the usual mechanical equilibrium equation $\partial \sigma_{\alpha z}^i / \partial z = f_\alpha^i(z)$ (see [33,75] and appendix I) with appropriate boundary conditions, that means $\sigma_{zz}^{i=l}(z = n_s + n_l) = 0$ at the free surface of the liquid (zero external pressure P is considered). Since there cannot be any excess of normal stress at the interfaces, $\sigma_{zz}^{i=s}(z=n_s)=\sigma_{zz}^{i=l}(z=n_s)$. Thus after a straightforward calculation there is:

$$
v_l^{1/3}\sigma_{zz}^l(z) = \left(e^{-z/\zeta} - e^{-n/\zeta}\right)\left[(2\gamma_l - \beta_{l/s})e^{n_s/\zeta} + (\beta_{l/s} - \beta_{l/S})\right] \quad for \quad n_s \le z \le n_s + n_l = n \quad \textbf{(29')}
$$

$$
\begin{aligned}
v_s^{1/3}\sigma_{zz}^s(z) &= (2\gamma_s - \beta_{s/S})\left(e^{-z/\zeta} - e^{-n_s/\zeta}\right) + (2\gamma_s - \beta_{s/l})\left(1 - e^{-n_l/\zeta}\right)\left(1 - e^{-(n_s-z)/\zeta}\right) \\
&\quad + a\sigma_{zz}^l(z = n_s) \quad for \quad 0 \le z \le n_s
\end{aligned} \quad \textbf{(29'')}
$$

In (29'') one sees that there are two inhomogeneous and one homogeneous contributions to the solid stress. The homogeneous part (the last term) is the pressure the liquid exerts on the underlying solid layers.

## VI.1/ Coherent epitaxy

The corresponding strains are given by Hookes law [75] applied to the epitaxial coherent strained solid s:

$$
\varepsilon_{zz}^s(z) = \frac{(1+\nu)(1-2\nu)}{E(1-\nu)}\sigma_{zz}^s(z) - \frac{2\nu}{1-\nu}m \quad \text{and} \quad \varepsilon_{xx}^s(z) = \varepsilon_{yy}^s(z) = m \quad \textbf{( 30)}
$$

Due to the coherency, in the solid therefore develops an in-plane stress:

$$
\sigma_{xx}^s(z) = \sigma_{yy}^s(z) = \frac{E}{1-\nu}m + \frac{\nu}{1-\nu}\sigma_{zz}^s(z) \quad \textbf{(31)}
$$

At the same there is in the liquid:

$$\varepsilon_{zz}^{l}(z) = \frac{\chi_l}{3}\sigma_{zz}^{l}(z) \quad \text{and} \quad \varepsilon_{xx}^{l}(z) = \varepsilon_{yy}^{l}(z) = 0 \qquad (32)$$

where $\chi_l$ is the volumic compressibility of the liquid.

Most easy to discuss is the case in absence of any liquid ($n_l$=0). Indeed $\sigma_{zz}^{s}(z)$ contains only the first term of (29") so that the strain along z according to (30) can be separated in two contributions a homogeneous one and a inhomogeneous one $\varepsilon_{zz}^{s}(z) = \varepsilon_{zz}^{s,in}(z) + \varepsilon_{zz}^{s,\mathrm{hom}}(z)$ with:

$$\begin{cases} \varepsilon_{zz}^{s,in}(z) = \dfrac{(1+\nu)(1-2\nu)}{Ea(1-\nu)}\left(2\gamma_s - \beta_{s/S}\right)\left(e^{-z/\zeta} - e^{-n_s/\zeta}\right)a \qquad for \qquad 0 < z < n_s \\[12pt] \qquad \varepsilon_{zz}^{s,\mathrm{hom}}(z) = -\dfrac{2\nu}{1-\nu}m; \qquad \varepsilon_{xx}^{s,\mathrm{hom}} = \varepsilon_{yy}^{s,\mathrm{hom}} = m \end{cases} \qquad (33)$$

Thus if the solid wets the substrate, $2\gamma_s - \beta_{s/S} < 0$, $\varepsilon_{zz}^{s,in}(z)$ is negative so that obviously the attraction field of the s/S interface contracts inhomogeneously the solid layers s but mostly the substrate nearest ones (we rule out the case $2\gamma_s - \beta_{s/S} > 0$ since then a s/S slab is no more stable but a 3D Volmer-Weber topology takes place [76-78]).

If there is some liquid on top of s then $\varepsilon_{zz}^{s}(z)$ is changed (see (30), (29)), mainly by we call the *field induced pressure of the liquid on the solid* represented by the last term of (29'') say $\sigma_{zz}^{l}(n_s)$ and explicitly from (29'):

$$\sigma_{zz}^{l}(n_s) = \left[\left(2\gamma_l - \beta_{l/s}\right) + e^{-n_s/\zeta}\left(\beta_{l/s} - \beta_{l/S}\right)\right]\left(1 - e^{-n_l/\zeta}\right)v_l^{-1/3} \qquad (34)$$

This pressure on top of the solid s P=-$\sigma_{zz}(n_s)$ is increasing with $n_l$. It is leaded by the $\Phi = 2\gamma_l - \beta_{ls} < 0$ wetting term when the substrate is far away. This positive pressure is however changed by the differential adhesion energy $\beta_{l/s} - \beta_{l/S}$ term when the solid s becomes thin.

## VI.2/ Coherent epitaxy CH$_4$(001)/MgO(100) (refer also to III4).

Let us illustrate these in-homogeneities with an accessible experimental example. Authors [22] measured also by neutron diffraction the mean lattice spacing (002) of the CH$_4$ (CD$_4$ for higher contrast) solid film on MgO substrate.

At 50 K *the film is dry* ($n_l$=0) with $n_s$= 5±0.15 and a mean inter-plane lattice parameter<$d_{2D}$>=2.85±0.03 Å is measured. At the same temperature bulk CH$_4$ has a (002) spacing (measured on the same sample) $d_{3D}$=2.98±0.01 Å. The mean lagrangian deformation is $\langle\varepsilon_{zz}\rangle_{exp.}=(\langle d_{2D}\rangle-d_{3D})/d_{3D}=-(4.3\pm1.3)10^{-2}$ which is clearly a measurable compression. Notice that at 50 K there is zero misfit strain (see III4) so that this value is the genuine inhomogeneous mean strain. In principle we can deduce from this value over $n_s$ layers with (30) the value: $2\gamma_s-\beta_{s/S}$ with measurements of a much better precision, let us say $10^{-3}$. Here let us use other experimental and theoretical data for $2\gamma_s-\beta_{s/S}$ and calculate $\langle\varepsilon_{zz}\rangle_{calc.}$. First from [43,44] rule reconsidered in appendix VII at $T_m=90$ K one has $\gamma_s=1.18\gamma_l=22.2\pm0.2$ erg cm$^{-2}$ where $\gamma_l=18.87$ erg cm$^{-2}$ comes from measurements [60]. Then from adsorption isotherms $CH_4$/MgO in the first solid monolayer there is the isosteric heats of adsorption $q_s=148\pm10$ meV [57] and $q_s=158\pm2$ meV [79]. The calculated value considering $CH_4$ as a hindered rotator at 90K [80] gives $q_s=152$ meV with a mean field lateral imbedding energy $\omega=22$ meV [80]. The adsorption energy of the single molecule therefore is $q_{ad}=130$ meV. In our scheme the bonding energy for the molecules closest to the substrate there is $q_{ad}=\beta_{s/S}a^2(1-e^{-1/\zeta})$ so that we can deduce $\beta_{s/S}=197\pm10$ ergcm$^{-2}$ (1 meV=1.6 $10^{-15}$ erg) therefore $2\gamma_s-\beta_{s/S}=-152\pm10$ erg.cm$^{-2}$. From (30) and the necessary elastic data in III4 then we calculate for $n_s=5$ the quantity $\langle\varepsilon_{zz}\rangle_{calc.}=-(3.5\pm0.4)10^{-2}$. Such a negative value corresponds to a compression as shown by experiments but with a smaller value likely due to both the experimental and thermoelastic data uncertainties. The unknown adhesion energies can then be obtained from the collected data in III.4, $\Gamma=(2\gamma_s-\beta_{s/S})-(\beta_{l/s}-\beta_{l/S})=-10.6\pm1$ erg.cm$^{-2}$ , $\Phi=2\gamma_l-\beta_{l/s}=-37\pm0.2$ ergcm$^{-2}$ (see III3). There is

$\beta_{l/s}=41.4\pm0.2$ erg.cm$^{-2}$ ;    $\beta_{l/S}=183\pm10$ erg.cm$^{-2}$    and    $\beta_{s/S}=197\pm10$ erg.cm$^{-2}$

We see that all adhesion energies are positive as it should be. That one of the liquid on its own solid is much smaller than hetero-adhesion l/S and s/S due to the strong "chemical binding" with MgO. Furthermore as a general rule for simple systems the adhesion energy of liquid (here $CH_4$) on any clean substrates should be smaller than the coherent adhesion of its solid on this substrate (here (001) $CH_4$ on (001) MgO) so that, $\beta_{l/S}<\beta_{s/S}$. Indeed across a l/S interface compared to the s/S one there is not only a small deficit of bond energy but essentially there are many supplementary possible configurations (entropy). [43-45,49,50]. The so-obtained values satisfy this inequality but with some uncertainties due to the error bars.

At 89.5 K according to data [21] the solid *$CH_4$ film becomes wetted*. On $n_s=3.8\pm0.15$ solid layers there are $n_l=4.4\pm0.2$ liquid layers and there was measured $<d_{2D}>=2.90\pm0.03$ Å very similar to the foregoing dry case. At that temperature the bulk (002) spacing of $CH_4$ is $d_{3D}=3.02\pm0.01$ Å so that the mean strain $<\varepsilon_{zz}>_{exp}=-(4.0\pm1.1)\%$ is similar to the previous one (dry case) but including the homogeneous $\varepsilon_{zz}$ component of the misfit is $\varepsilon_{zz}^{hom}=-\dfrac{2\nu}{1-\nu}m=+(2.0\pm0.15)\%$. That means the measured part devoided of the misfit $\left\langle\varepsilon_{zz}\right\rangle'_{exp}=-(6.0\pm0.1)\%$ which is quite greater than in the dry case.

It is instructive to look at the different components of this strain. For this purpose let us calculate the field pressure the liquid exerts on the top of the solid film using (34). One finds $\sigma_{zz}^{s}(n_s=3.8, n_l=4.4)\approx-(1.1\pm0.2)\ 10^{8}$ dynes cm$^{-2}$ that means about P=+10$^{2}$ bars. Doubling the liquid thickness would increase this pressure by only 1%, however decreasing the number of solid layers by a factor 2 would increase this pressure by a factor 3 via the differential adhesion in (34) which gains a more important negative value. This illustrates how closely to the interfaces l/s and s/S are localised the induced inhomogeneities for $\zeta\approx1$.

Finally calculating the mean strain with (30) and (29', 29'', 11, 34), and the formerly obtained data excluding also the contribution of the in-plane misfit strain in (30) one finds for $n_s=3.8$ and $n_l=4.4$; $\left\langle\varepsilon_{zz}\right\rangle'_{calc}=\left[-3.5+0.08-0.7\right]10^{-2}=-(4.1\pm0.7).10^{-2}$

This value of the wet case is also a contraction but greater than in the dry case. However it is smaller than the measured one (-6.0±0.1)%. There may be a systematic error but we can't tell where it is residing in the measurements, the calculations or both.

Let us note that the sequence of numbers in the expression of $\left\langle\varepsilon_{zz}^{s}\right\rangle'_{calc}$ corresponds to the successive factors of (29''). The last one is due to the field pressure in the liquid. The two first terms are the inhomogeneous contributions. The leading term is due to the substrate field characterized by $2\gamma_s-\beta_{s/S}$ whereas the smallest term is due to $\widetilde{\Phi}=2\gamma_s-\beta_{s/l}=1\pm0.2$ erg.cm$^{-2}$, the non wetting factor of the liquid by its solid. Let us stress that even small but positive, this term in fact guarantees that solid $CH_4$ does premelt at T<$T_m$ at the interface s/v and not in the interface s/S since $\Phi<0<\widetilde{\Phi}$.(Such an interfacial premelting could in principle happen for $\widetilde{\Phi}<\Phi<0$ (see III2)).

Figure 12 as an illustration of the half wetted system $n_s=n_l=4$ gives the inhomogeneous strain $\varepsilon_{zz}$ in the solid and the supernatant liquid (ingredients of the calculations are those of $CH_4$/MgO but with zero misfit m=0). At the solid/liquid interface (z=4) there is a strain discontinuity. Obviously to this inhomogeneous compressive contribution (because $2\gamma_s - \beta_{s/S} < 0$) must be added the homogeneous contribution of the misfit m that can be positive or negative according to the system (for $CH_4/MgO$, m=-1.5% at 90 K).

**VI3/ Inhomogeneity effect on melting curves.**

Using (30) (29') and (29'') for strain and stress in the film it is easy to write down the elastic energy of the deposited material A ($n_s + n_l$) per unit surface of S

$$W_{el} = \frac{1}{2b^2} \int \sigma_{\alpha\beta}^i \varepsilon_{\alpha\beta}^i dV = \frac{Eam^2}{1-\nu} n_s + \frac{Y_s a}{2} \int_0^{n_s} \left[\sigma_{zz}^s(z)\right]^2 dz - \frac{\chi_l v_l^{1/3}}{6} \int_{n=n_s+n_l}^{n_s} \left[\sigma_{zz}^l(z)\right]^2 \tag{35}$$

with $Y_s = \dfrac{(1+\nu)(1-2\nu)}{(1-\nu)E}$ and $\dfrac{\chi_l}{3}$ respectively the linear compressibility of solid and liquid.

The inhomogeneous distributions of normal stress in liquid and solid in the integrals are given by (29') and (29''). However we must remember that $\sigma_{zz}^s(z)$ contains a homogeneous contribution (see last term of (29'')). The situation is complex however so that the effect on the melting curve will be only numerically treated. In figure 13 we consider the case of $CH_4/MgO$ where $T_m$=90 K and n=8. In the upper insert of figure 13 are shown the misfit strain energy and the total elastic energy of the film ($n_s + n_l$) calculated at 90 K. Obviously the misfit energy (dotted curve) is a straight line whereas due to the inhomogeneity contribution the total elastic energy is quite concentrated in the first two layers of the solid s. The contribution of the elastic energy of the liquid is so small that it can be neglected. However the field induced pressure (34) exerted by the liquid l on the solid s contributes to the total energy. It is given by the difference in between the linear parts of the total and the misfit energy.

The right side insert of figure 13 gives the melting temperature shift along the melting curve due to the elastic energy in the solid s and the melting temperature shift due to the misfit energy. For the dry solid ($n_s$=8, $n_l$=0) only the misfit energy contributes to the shift, whereas because of the strong inhomogeneity of the total elastic energy the temperature shift strongly deviates from the usual one when only a few solid layers remain on the substrate S. For the example of $CH_4/MgO$ there is an exact compensation and thus no more temperature shift for $n_s$=3.

Finally let us consider the inhomogeneity effect on the melting curve. In the main part of figure 13 the full curve is that one of figure 8 where inhomogeneity effects have not be taken into account. The dotted curve is that one calculated when inhomogeneity effects are taken into account. It is important to *note that elastic inhomogeneities only distort the final part of the melting curve when only a few solid layers resist to melting*. This behaviour only described for the CH$_4$/MgO system is general for coherent astride premelting ($\Phi$<0, $\Gamma$<0): the misfit energy shifts the melting curve towards lower temperature (see II1) and inhomogeneous strain due to the interfacial fields distorts the melting curve at temperature higher than $T_m^{'}$ towards higher temperature. In the case of coherent boosted premelting ($\Phi$<0, $\Gamma$>0) there can be no detectable effect of the field inhomogeneity since all the melting curve is located at T<$T_m^{'}$.

Incidentally let us remark that for coherent epitaxies the maximum number of equilibrium solid layers is given at undersaturation by the solid/solid wetting factor $2\gamma_s - \beta_{s/S}$ <0, see [81,82]

$$n_s / \zeta = \ln\left(\frac{|2\gamma_s - \beta_{s/S}|}{E\xi am^2 / (1-\nu)}\right) \qquad \textbf{(36)}$$

For the above treated case CH$_4$/MgO (VI2) where we earned the values $\zeta$=1 and $2\gamma_s - \beta_{s/S}$ =-152 ergcm$^{-2}$ there is 6<n$_s$<7 which is quite in agreement with the limiting numbers of steps observed in the adsorption isotherms [57, 58, 79]. Beyond this number of layers 3D Stranski-Krastanov crystals may build up.

Finally *for glissile epitaxies* where surface stress produces an homogeneous in-plane stress of the solid s motivating in-plane and normal to plane strains, regardless of natural misfit as stated in V, body forces we discussed in this section produce of course supplementary z-inhomogeneities and of course a small in-plane strain contribution. The effect on the melting curve is given by the same terms of (35) excepted the first term has to be replaced by (24) (20). Thus to the distorsion of the melting curve due to the inhomogeneity has to be added the distorsion due to the surface stress. As for coherent epitaxy we treated here in detail, these effects can be neglected at T≤T$_m$ along the melting curve but play some role at T>T$_m$ that means the melting of the last solid layers..

## VII/ Conclusion and Out-look

We have treated the surface melting of epitaxial films that means surface melting of *a finite size solid* A overgrowing in a regular way onto a lattice mismatched substrate B. We have shown that finite size and epitaxial contact both lead to new surface premelting properties different from the well-known surface premelting properties of semi-infinite solids.

*The size effects* can be simply taken into account in a macroscopic approach of surface melting in which usual surface and interfacial excess quantities are expressed in terms of short range chemical interactions duly extended by longer range interactions. Apart the usual coefficient $\Phi<0$ (wetting factor) characterising also the interaction between the v/l and l/s interfaces, a new coefficient $\Gamma$ describing the interactions in between l/s and s/S interfaces introduces naturally. According to the sign of $\Gamma$ two types of surface melting can be predicted. They are independant of the type of long range interactions we consider, as illustrated with screened interactions or Van der Waals interactions. When $\Gamma>0$ the l/s-s/S interfaces attraction boost the premelting which then occurs in two steps: a continuous premelting followed by a first order transition. When $\Gamma<0$ the repulsion of these interfaces refrains the premelting so that a part of the deposited solid remains solid above its melting point. A very serious lack of interfacial thermodynamical data limits the predictive effectiveness of the theory. However careful analysis of experimental melting curves with other data from separate analytical tools allows to extract these data. We could illustrate this with the $CH_4/MgO$ system where neutron and electronic diffraction, inelastic neutron scattering and adsorption isotherms have been used. This system belongs to the astride type where premelting and overheating occurs. All interfacial energies could be determined.

Because of *epitaxial contact* a new ingredient, the elastic energy stored in the epitaxial layer has to be introduced. It can be divided in two contributions: a bulk contribution and a surface contribution.

*The bulk contribution* essentially originates from the lattice mismatch between the deposit and its substrate. Obviously it is all the more important the epitaxy is coherent and the misfit high. The homogeneous elastic energy so induced only shifts at constant misfit the melting curve to lower temperature. However, to this homogeneous contribution adds an inhomogeneous bulk elastic energy due to the interaction in between the bulk of the various material layers (another type of size effects). Though measurable (and measured in the $CH_4/MgO$ example) this localised strain only produces some distortion in the final part of the melting curve.

*The surface contribution* originates from the intrinsic surface and interfacial stresses considered as excess quantities and, coupled with bulk strain, modifies $\Phi$ and $\Gamma$. Obviously this contribution is all the more important the misfit is great in the case of coherent epitaxy. In the experimental case of $CH_4/MgO$, due to the small and even vanishing misfit (according to temperature), this effect is hardly measurable. This contribution is however dominant for *glissile epitaxies* where strain is induced by the surface stress itself. It distorts the initial part of the melting curves. As for the interfacial energy the lack of surface and interfacial stress data limits effective predictions of the theory for case studies. We could show that the system Kr/graphite approached by heat capacity studies, enters in this category of epitaxies. The $\Phi, \Gamma < 0$ data could be extracted as well as the interfacial stress $s_{sS}$ of the Kr/graphite interface. Even not yet measured directly by experiments it is possible to calculate the in-plane strain in the successive Kr layers nearest the graphite substrate.

We could not find in literature studies of surface melting of thin metal films on metals, metals films on compounds (oxides) or organic crystals on any stable substrate. *This is an unexplored field where surface melting studies could bring* not only the $\Phi$ and $\Gamma$ (interfacial energies) data from the melting curves or their strain derivatives that means interfacial stresses. Furthermore may be that the case of boosted premelting could be find in the future on systems where $\Gamma > 0$.

We have shown how useful and, up to now, unknown data can be obtained provided that these measurements are connected with careful surface studies with X-ray diffraction and scattering needing in these cases the use of synchrotron radiation, adsorption measurements with Auger or Mass Spectrometry. To these essential auxiliary studies we added at the end of section III2 some proposals to determine interfacial energy or adhesion energy by measuring isothermal transfer of wetting layers. The proposals would hold too for the measurement of interfacial stress by the method of curvature [62, 64]. Both measurements, energy and stress may be done simultaneously in the same isothermal vessel. Let us recall that due to the various influences of elastic bulk and interfacial elastic energies we showed, that candidates of couples s/S for surface melting studies have first to be characterised fully as epitaxial couples for several thicknesses and temperatures. The bulk phases, deposit and substrate, have to be stable, without any intermixing. (Our predictions are not valid when alloying occurs). Furthermore the surface and interfaces have to be stable intrinsically (orientations along cusps of the γ-plot) during their elaboration and during temperature raise. Inelastic relaxations when occurring spoil or change the melting curves. Are concerned dislocation introduction and Stranski-Krastanov transitions for coherent epitaxies. Let us recall that real glissile epitaxies are not subject to these relaxations up to great thicknesses due to their non increasing strain energy.. Their practical interest is clear for these reasons however up to now this field has not been really explored. In electronic applications, where the activity is the more important, only degenerate epitaxies are used yet (see foot note 3) excepted some brilliant examples as MnAs/GaAs where the prism face of hexagonal MnAs meets the cube face of GaAs [83] and where in the contact plane there is a multiple coincidence lattice. Metals on oxides system belong to the same type of non degenerated epitaxies [84] where perfect films can be grown and may be good candidates for surface melting studies of nanoscopic films.

## Appendix I

## General equilibrium conditions of stressed solids in contact with a fluid

The general equilibrium conditions of fluid/solid systems have been established by Gibbs [33] then became tractable by Larché and Cahn [34]. The equilibrium conditions so obtained by variational calculus can be splitted in thermal, mechanical and chemical equilibria and associated boundaries conditions. For a stressed solid s (stress tensor $\sigma_{ij}$) in equilibrium with a fluid (hydrostatic pressure P) the mechanical and chemical equilibrium reads successively:

* $\partial \sigma_{ij} / \partial x_j = 0$ (in absence of any body forces as gravity or others) with $\sigma_{ij} n_j = -P n_j \delta_{ij}$ at the solid/fluid interface characterised by the normal vector $n_i$. Of course we neglect gravity forces, mostly negligible for thin films (see the end of III2). However in appendix VI we consider the body forces induced in s and l by the finite size effect.

* $U^s - TS^s + PV^s = \mu^f N^s$ where $N_s$ is the number of moles *transferred* from the solid to the fluid, $U^s$ and $S^s$ the internal energy and entropy density of the solid and $V^s$ its molar volume. $\mu^f$ is the chemical potential of the fluid. Let us note that Gibbs avoids to define a chemical potential for one component solids. Further discussion can be found in [34]. The previous chemical equilibrium condition can also be written

$$G^s \left( T, \sigma_{ij} \right) = \mu^f N^s \qquad \textbf{(I)}$$

Where $G^s \left( T, \sigma_{ij} \right)$ is the Gibbs free energy of the stressed solid at temperature T.

In absence of any stress-chemical interaction, what is obviously the case for a pure solid, the elastic energy behaves as an excess energy [34] so that the Gibbs energy (I) of the solid in equilibrium with a fluid may be written in the framework of linear elasticity:

$$G^s (T, \sigma_{ij}) = G^s (T,0) + \frac{1}{2} S_{ijkl} \sigma_{ij} \sigma_{kl} V^s = \mu^f N^s \qquad \textbf{(II)}$$

where $S_{ijkl}$ are the usual elastic compliances and $V^s$ the molar volume of the solid under zero stress.

At the same the Gibbs free energy of a fluid (hydrostatic pressure P) in equilibrium with a solid can be written

$$G^f (T,P) = G^f (T,0) - \frac{1}{2} \chi P^2 V^f = \mu^f N^f \qquad \textbf{(III)}$$

where now $N^f$ is the number of moles of liquid *transferred* to the solid and $\chi$ is the compressibility of the fluid, $V^f$ its molar volume at zero pressure. Let us note that when capillary effects are neglected the Gibbs energy change due to the solid/fluid transformation reads $\Delta G = G^f(T,P) - G^s(T,\sigma_{ij}) = \mu^f(N^f - N^s)$. Thus the solid melts when $N^f - N^s < 0$ that means for $\Delta G < 0$. The equilibrium state is thus described by $\Delta G = 0$, and G thus is the appropriate thermodynamic potential for studying the solid/fluid transition under stress.

## Appendix II

### Mechanical equilibrium of homogeneously strained slab

In the case of a biaxially strained cubic solid slab (001) ($\varepsilon_{11} = \varepsilon_{22} = m$) covered by a fluid characterised by its hydrostatic pressure P the strain and stress tensors of the solid s read:

$$[\varepsilon] = \begin{pmatrix} m & 0 & 0 \\ 0 & m & 0 \\ 0 & 0 & \varepsilon_{33} \end{pmatrix} \text{ and } [\sigma] = \begin{pmatrix} \sigma_{11} & 0 & 0 \\ 0 & \sigma_{11} & 0 \\ 0 & 0 & \sigma_{33} \end{pmatrix}$$ where owing to mechanical equilibrium

conditions (see appendix I) at the fluid/solid interface there is $\sigma_{33} = -P$. No body forces are considered (see remark in appendix I). The unknown data $\varepsilon_{33}$ and $\sigma_{11}$ may be obtained from Hooke's law [75]: $\varepsilon_{11} = \frac{1}{E}[\sigma_{11} - \nu(\sigma_{11} + \sigma_{33})]$ and $\varepsilon_{33} = \frac{1}{E}[\sigma_{33} - \nu(\sigma_{11} + \sigma_{11})]$ with E the Young modulus and $\nu$ the Poisson ratio in the (001) plane[8]. Thus using $\varepsilon_{11} = m$ and $\sigma_{33} = -P$ one obtains:

$\varepsilon_{33} = -\frac{2\nu m}{1-\nu} - \frac{P}{E}\frac{1-\nu-2\nu^2}{1-\nu}$ and $\sigma_{11} = \frac{Em - \nu P}{1-\nu}$. Thus the free elastic energy density $\frac{1}{2}\sigma_{ij}\varepsilon_{ij}$ reads:

$$W_{el} = \frac{Em^2}{1-\nu} + \frac{P^2}{2E}\frac{(1+\nu)(1-2\nu)}{1-\nu} \qquad \textbf{(I)}$$

---

[8] The biaxial modulus for this (001) slab reads $\left.\frac{E}{1-\nu}\right|_{(100)} = C_{11} + C_{12} - 2\frac{C_{12}^2}{C_{11}}$ ; $\nu|_{(100)} = \frac{C_{12}}{C_{11}+C_{12}}$, $C_{ij}$ being the elastic moduli. For a (111) slab on a (111) surface with $\varepsilon_{//} = m$ there is to substitute above $\left.\frac{E}{1-\nu}\right|_{(111)} = \frac{6(C_{11}+2C_{12})C_{44}}{C_{11}+2C_{12}+4C_{44}}$ ; $\nu|_{(111)} = \frac{C_{11}+2C_{12}-2C_{44}}{2(C_{11}+C_{12}+C_{44})}$ [85]. Notice that for cubic crystals the elastic moduli $C_{ij}$ are connected to the compliances moduli $S_{ij}$ by $C_{11} = (S_{11}+S_{12})/(S_{11}-S_{12})(S_{11}+2S_{12})$ , $C_{12} = -S_{12}/(S_{11}-S_{12})(S_{11}+2S_{12})$ and $C_{44} = 1/S_{44}$.

At the same the strained volume

$$V_{def}^s = V^s\left(1+\varepsilon_{ii}\right) = V^s\left(1+2m\frac{1-3\nu}{1-\nu}-\frac{P}{E}\frac{(1+\nu)(1-2\nu)}{1-\nu}\right) \qquad \textbf{(II)}$$

where $V^s$ is the undeformed molar volume of the solid. One distinguishes clearly in I and II the effect of misfit strain m and that of hydrostatic pressure P where there are no cross terms. Notice that when units of E are in ergcm$^{-3}$ and $10^{10}<E<10^{12}$ as usually, since 1bar=1.033 Atm=$10^6$ ergcm$^{-3}$, there is for P=1bar P/E=$10^{-4}$-$10^{-6}$ a very small correction to all quantities $\varepsilon_{33}$, $\sigma_{//}$, I and II.

## Appendix III

## Surface energy, adhesion energy and interfacial energy of finite size planar slabs

Consider not only short range "chemical interactions" but also longer ranged ones as -r$^{-6}$ dispersion forces or $-\frac{1}{r}e^{-r/\zeta_a}$ for screened Coulomb forces [86]. To the chemical bonding between the layers of a slab of thickness d there adds a d$^{-3}$ or a $e^{-d/\zeta_i}$ contribution. A slab of $n_i$ layers squeezed between two semi-infinite parts of the same matter i and extracted from there has thus an excess energy equal to twice the surface energy[9] $\gamma_i$ of the finite slab and which reads for example for screened Coulomb forces

$$\gamma_i\left(n_i\right) = K_i\sum_{n=0}^{n_i-1}e^{-n/\zeta_i} = K_i\frac{1-e^{-n_i/\zeta_i}}{1-e^{-1/\zeta_i}}$$

The material constant $K_i$ characterising the chemical interactions can be obtained from asymptotic considerations since when the slab becomes very thick its excess energy has to tend towards the usual surface energy $\gamma_i^\infty$. Thus there is $K_i/\left(1-e^{-1/\zeta_i}\right) = \gamma_i^\infty$ so that the specific surface energy of a thin slab of material i ($n_i$ layers) reads [87,82,67]

$$\gamma_i\left(n_i\right) = \gamma_i^\infty\left(1-e^{-n_i/\zeta_i}\right) \qquad \textbf{(I)}$$

For r$^{-6}$ dispersion forces one obtains similarly

$$\gamma_i\left(n_i\right) = \frac{K_i}{2}\sum_1^{n_i}n^{-3} = \alpha\gamma_i^\infty\sum_1^{n_i-1}n^{-3} \qquad \text{with } 1/\alpha = \sum_1^\infty n^{-3} = 1.202..$$

A similar expression has been obtained for the adhesion energy $\beta_{i/j}$ ( the work for separating a body i from a body j producing an unit area of i and j) of $n_i$ layers of material i coherently bond on $n_j$ layers of material j [67]

$$\beta_{i/j}\left(n_i, n_j\right) = \beta_{i/j}^{\infty}\left(1 - e^{-n_i/\zeta_i}\right)\left(1 - e^{-n_j/\zeta_j}\right) \tag{II}$$

Notice that the interfacial energy $\gamma_{ij}\left(n_i, n_j\right)$ between the slab i and the slab j cannot be calculated directly but has to be deduced from surface energies $\gamma_i\left(n_i\right)$ and adhesion energy $\beta_{i/j}\left(n_i, n_j\right)$ by the Dupré relation [89]

$$\gamma_{ij}\left(n_i, n_j\right) = \gamma_i\left(n_i\right) + \gamma_j\left(n_j\right) - \beta_{i/j}\left(n_i, n_j\right) \tag{III}$$

obtained by means of an energy cycle [78].

For our purpose (see figure 2a where i=l, j=s and k=S) it is necessary to calculate the adhesion energy $\beta_{i/jk}$ of a material i ($n_i$ layers) over a *composite slab* constituted by $n_j$ layers of material j over $n_k$ layers of material k noted $i(n_i)/j(n_j)/k(n_k)$. Such an adhesion energy can be obtained by a same straightforward summation procedure. Nevertheless it can also be obtained more easily by means of a thermodynamic process where the 3-composite slab $i(n_i)/j(n_j)k(n_k)$ is obtained as a combination of single 2-composite slabs as:

$$i(n_i)/j(n_j)k(n_k) = i(n_i)/j(n_j) + i(n_i + n_j)/k(n_k) - i(n_j)/k(n_k)$$

with relations of type II valid for binary slabs.

Thus there is

$$\beta_{i/jk}(n_i, n_j, n_k) = \beta_{i/j}^{\infty}\left(1 - e^{-n_i/\zeta_i}\right)\left(1 - e^{-n_j/\zeta_j}\right) + \beta_{i/k}^{\infty}\left(1 - e^{-(n_i/\zeta_i + n_j/\zeta_j)}\right)\left(1 - e^{-n_k/\zeta_k}\right) - \beta_{i/k}^{\infty}\left(1 - e^{-n_j/\zeta_j}\right)\left(1 - e^{-n_k/\zeta_k}\right)$$

This expression can be rearranged and thus reads:

$$\beta_{i/jk}(n_i, n_j, n_k) = \beta_{i/j}^{\infty}\left(1 - e^{-n_i/\zeta_i}\right)\left(1 - e^{-n_j/\zeta_j}\right) + \beta_{i/k}^{\infty} e^{-n_j/\zeta_j}\left(1 - e^{-n_i/\zeta_i}\right)\left(1 - e^{-n_k/\zeta_k}\right) \tag{IV}$$

Notice that relation IV is written for the sequence i/jk so that the total surface and interfacial free enthalpies written in the same sequence read for the 3 composite slab:

$$g^{surf}(n_i, n_j, n_k) = \gamma_i(n_i) + \gamma_{i/jk}(n_i, n_j, n_k) + \gamma_{j/k}(n_j, n_k) + \gamma_k(n_k) \tag{V}$$

where $\gamma_{i/jk}\left(n_i, n_j, n_k\right)$ is the interfacial energy of material i ($n_i$ layers) onto the composite material $j(n_j)/k(n_k)$ and $\gamma_{j/k}\left(n_j, n_k\right)$ the interfacial energy of j ($n_j$) onto k ($n_k$).

Using Dupré's equation III under the form

$$\gamma_{i/jk}(n_i, n_j, n_k) = \gamma_i(n_i) + \gamma_j(n_j) - \beta_{i/jk}(n_i, n_j, n_k) \tag{VI}$$

$$\gamma_{j/k}(n_j, n_k) = \gamma_j(n_j) + \gamma_k(n_k) - \beta_{j/k}(n_j, n_k) \tag{VII}$$

and relations of type IV, II and I one obtains for the total free enthalpy (V) of the 3-component slab:

---

[9] When there is only one index $\gamma_i$ it means the interfacial energy of i in respect to v (its vapour or vacuum)

$$g^{surf}(n_i, n_j, n_k) = \left(2\gamma_i^\infty - \beta_{i/j}^\infty\right)\left(1 - e^{-n_i/\zeta_i}\right) + \left(2\gamma_j^\infty - \beta_{j/k}^\infty\right)\left(1 - e^{-n_j/\zeta_j}\right) +$$
$$\beta_{i/j}^\infty e^{-n_j/\zeta_j}\left(1 - e^{-n_i/\zeta_i}\right) + \beta_{j/k}^\infty e^{-n_k/\zeta_k}\left(1 - e^{-n_j/\zeta_j}\right) \qquad \textbf{(VIII)}$$
$$-\beta_{i/k}^\infty e^{-n_j/\zeta_j}\left(1 - e^{-n_i/\zeta_i}\right)\left(1 - e^{-n_k/\zeta_k}\right) + 2\gamma_k^\infty\left(1 - e^{-n_k/\zeta_k}\right)$$

Finally for the system of figure 2a, identifying the phases l,s,S respectively by i=l, j=s, k=S and putting $n_k \rightarrow \infty$ there is

$$g^{surf}(n_l, n_s) = 2\gamma_S + \left(2\gamma_l - \beta_{l/s}\right)\left(1 - e^{-n_l/\zeta_l}\right) + \left(2\gamma_s - \beta_{s/S}\right)\left(1 - e^{-n_s/\zeta_s}\right) +$$
$$\left(\beta_{l/s} - \beta_{l/S}\right)e^{-n_s/\zeta_s}\left(1 - e^{-n_l/\zeta_l}\right) \qquad \textbf{(IX)}$$

We omitt in IX and the following the subscript $\infty$ but all the surface energies $\gamma_i$ and adhesion energies $\beta_{ij}$ are meant to be the usual macroscopic quantities. Relation IX is used in section II formula (7). Notice that when solid s becomes thick (fig 1a) then (IX) is only finite size dependent upon the liquid film $n_l/\zeta_l$ and formula IX reduces to:

$$g^{surf}(n_l, n_s = \infty) = \left(2\gamma_l - \beta_{ls}\right)\left(1 - e^{-n_l/\zeta_l}\right) + C^{te} \qquad \textbf{(X)}$$

A useful planar system is when i is a thick cover glass (cg) over the melt of thickness $n_j = n_l$ of the solid k=s of the semi infinite crystal $n_k = n_s \rightarrow \infty$. Making these transformations in VIII there is:

$$g^{surf}(n_l, n_s = \infty) = \left(2\gamma_{cg} - \beta_{l/cg}\right) + \left(2\gamma_l - \beta_{l/s}\right)\left(1 - e^{-n_l/\zeta_l}\right) + \beta_{l/cg}e^{-n_l/\zeta_l} - \beta_{s/cg}e^{-n_l/\zeta_l} + C^{te} \quad \textbf{(XI)}$$

Relation XI is used in section II1.1 formula (13').

The general relation (VIII) when i=k and $n_i = n_k \neq n_j$ gives also the surface free enthalpy of a non supported solid slab j=s of thickness $n_s$ covered on both faces by a liquid i=k=l of $n_l$ layers so that the total number of layers is $n = n_s + 2n_l$. Thus there is:

$$G^{surf}(n_l, n_s, n_l) = 4\gamma_l\left(1 - e^{-n_l/\zeta_l}\right) + 2\gamma_s\left(1 - e^{-n_s/\zeta_s}\right) - 2\beta_{l/s}\left(1 - e^{-n_l/\zeta_l}\right)\left(1 - e^{-n_s/\zeta_s}\right) - 2\gamma_l e^{-n_s/\zeta_s}\left(1 - e^{-n_l/\zeta_l}\right)^2 \textbf{(XII)}$$

Notice that when the solid vanishes $n_s \rightarrow 0$, $G^{surf}(n_l, 0, n_l) = 2\gamma_l\left(1 - e^{-n/\zeta_l}\right)$, when the slab is dry $n_l \rightarrow 0$ $G^{surf}(0, n_s, 0) = 2\gamma_s\left(1 - e^{-n/\zeta_s}\right)$ what is quite consistent. Relation XII has to be used for calculations of the type Sakaï [25] we mentioned in the introduction.

It may be that such a thin free standing slab is not easy to handle. Putting a thick cover glass on each face (XII) has to be changed by substituting only $\gamma_l$ by $\gamma_{cg/l}$, $\gamma_s$ by $\gamma_{cg/s}$ and adding $2\gamma_{cg}$.

# Appendix IV

## Crystallising back a finite layer of liquid on a substrate (fig 2a)

The activation barrier for crystallisation (we discuss in III1) at T=T$_l$ reads

$$\Delta G_{l\to s}(T = T_l) = G(T_l, n_s = n - n^{l,\max}, n_l = n^{l,\max}) - G(T_l, n_s = 0, n_l = n) \qquad \textbf{(I)}$$

where T$_l$, the temperature beyond which the material A is completely melted, is given by (14a) and $n^{l,\max}$ the number of liquid later at T=T$_l$ is given by (13). Incorporating then expressions (14a) and (13) in the expression of the activation barrier $\Delta G_{l\to s}(T = T_l)$ (i) yields:

$$\Delta G_{l\to s}(T = T_l) = N_{av}a^2\left[-\left(\frac{n}{2} + \ln\left(\frac{\Phi}{\Gamma}\right)\right)\left(\Gamma - \Phi e^{-n}\right) - \Phi e^{-n} + \frac{\Phi}{\Gamma}e^{-n/2} + (\beta_{l/s} - \beta_{l/S})f(n) - (2\gamma_s - \gamma_{sS})g(n)\right]$$

where $f(n) = 1 - e^{-n} - \frac{\Gamma}{\Phi}e^{-n/2} - \left(\frac{\Gamma}{\Phi}\right)^2 e^{-(n/2)^2}$ and $g(n) = 1 - \frac{\Gamma}{\Phi}e^{-n/2}$ and where furthermore we use $\Delta U_m = -T_m\Delta S_m$. Thus for great enough values of n so that $e^{-n} << 1$, $\frac{\Phi}{\Gamma}e^{-n/2} << 1$ and

$\frac{\Gamma}{\Phi}e^{-n/2} << 1$, there is:

$$\Delta G_{l\to s}(T = T_l) \propto N_{av}b^2\Gamma\left(\frac{n}{2} + 1\right) \qquad \textbf{(II)}$$

# Appendix V
## Surface energy of a solid and its melt of simple substances

Miedema et al. [43-45] have computed on many examples of simple pure metals a useful empirical relation between the mean surface energy of a solid and its liquid at zero K and Pluis [11] near the melting point. Surface energies of metals are anisotropic but only about 2-3% and surface energy of liquid are easily measured. We give here a general approximate derivation for non metals. Condensed phases i=l,s have surface specific energies $\gamma_{i,v}$ in respect to their very diluted vapour v defined according to Born and Stern [88] by $\gamma_{i,v} = W_{i,v}/2S_{iv}$ that means the reversible work of separation $W_{iv}$ along a planar surface $S_{iv}$ in two parts. This work represents also bonds to be broken so that $W_{iv} = k\Delta U_{iv}$ is proportional to the phase transition energy $\Delta U_{iv} = U_v - U_i$ of atoms leaving from i=s,l to the vapour v. Therefore $\gamma_s / \gamma_l = (\Delta U_{sv} / \Delta U_{lv})(V_l / V_s)^{2/3}$ where $V_l$ and $V_s$ are the molar volumes of the condensed phases s and l, $\Delta U_{sv}$ the sublimation energy of the solid and $\Delta U_{lv}$ the vaporisation energy of the liquid. Since one has $\Delta U_{sv} = \Delta U_{sl} + \Delta U_{lv} = \Delta U_m + \Delta U_{vap}$ where m and vap means melting and vaporisation, at a common temperature T=Tm (the melting temperature) one has:

$$\gamma_s / \gamma_l \big|_{T_m} = \left(1 + \Delta U_m(T_m) / \Delta U_{vap}(T_m)\right)(V_l / V_s)^{2/3}_{T_m} \qquad \textbf{(I)}$$

relation which contains in principle known bulk data.

With some less accuracy this relation may become a numerical rule. At the melting point $\Delta U_m(T_m) = T_m \Delta S_m(T_m)$ where positional melting entropy is according to the Matignon's rule $2 < \Delta S_m(T_m) = < 3$ cl mole$^{-1}$ and $\Delta U(T_b) = T_b \Delta S(T_b)$, with according to the Trouton's rule $\Delta S_{vap}(T_b) = 22$ cl mole$^{-1}$ at the boiling point $T_b$ (where the vapour pressure is 1 Atm). Since $\Delta U_{vap}(T_m) = \Delta U_{vap}(T_b) - \int_{T_m}^{T_b} \Delta c_{vap} dT$ with $\Delta c_{vap} = c_p^{vap} - c_p^s \approx -R/2$ in the high temperature limit, even for $T_m < T_b < 2T_m$ taken as extrema this correction only amounts to less than 4% and can be neglected. Simple usual substances increase their volume by 5-10% when melting. From these extrema one has with (I)

$$\gamma_s / \gamma_l \big|_{T_m} = \left(1 + T_m \Delta S_m(T_m) / T_b \Delta S_{vap}(T_m)\right)(V_l / V_s)^{2/3} \qquad \textbf{(II)}$$

or numerically $1.09 < \gamma_s / \gamma_l < 1.17$.

For explicit calculations and when data are available we will use relation (II). For example for $CH_4$ at $T_m = 90$ K there is from [60] $\Delta S_m(T_m) = 2.48$, $\Delta S_{vap}(T_m) = 17.4$ cl mole$^{-1}$, $T_b = 112$ K, $V_l(T_m) = 36.5$ and $V_s(T_m) = 32.4$ cm$^3$ mole$^{-1}$ so that $\gamma_s / \gamma_l \big|_{T_m} = 1.18$.

For Kr at $T_m = 116$ K there is from [60] $\Delta S_m(T_m) = 3.37$, $\Delta S_{vap}(T_m) = 18.01$ cl. mole$^{-1}$, $T_b = 120$ K, $V_l(T_m) = 34.7$ and $V_s(T_m) = 29.8$ cm$^3$ mole$^{-1}$ so that $\gamma_s / \gamma_l \big|_{T_m} = 1.30$.

# Appendix VI

## Shuttleworth' relations and reference state

The work necessary to create a surface (area $A_o$) of a material A (of surface energy $\gamma_A$ and surface stress $s_A$) then to deform this surface from $A_o$ to A reads $\Delta F = \gamma_A A_o + s_A(A - A_o)$. Defining then the isotropic in-plane strain $\varepsilon$ by $A = A_o(1 + \varepsilon)^2$, one obtains $\Delta F = (\gamma_A + 2\varepsilon s_A)A_o \approx (\gamma_A + 2\varepsilon(s_A - \gamma_A))A$. The first expression of $\Delta F$ is said to be written in Lagrangian coordinates that means in the non-deformed reference state. Within this reference state the Shuttleworth relation reads $\partial\gamma_i / \partial\varepsilon = s_i$. The second expression is written in Eulerian coordinates that means in the deformed reference state. In this case the Shuttleworth relation reads $\partial\gamma_i / \partial\varepsilon = s_i - \gamma_i$. (see [63,74] for more details).

# Appendix VII
## Inhomogeneous body fields in the composite slab

The excess field felt by a layer in a composite material (l/s/S) in respect to the same layer in a continuous material can be obtained from the difference of interactions between a layer located at the distance z from the substrate S on which is deposited the composite slab (s/l) and a equivalent layer located at the same level but in a continuous material. Since the so-called z-layer can be in the liquid or in the solid part of the composite slab, there are two excess fields $\Omega_l(z)$ and $\Omega_s(z)$ according to the nature (liquid or solid) of the z-layer.

These long range interactions between a layer and the whole of the material can be calculated as in appendix III by summation of screened Coulomb or van der Waals forces. Nevertheless we have to distinguish two cases according to the location of the z-layer.

- For a liquid layer ($z>n_s$) the summation of all the involved interactions reads (with $\zeta_s=\zeta_l=\zeta$):

$$\omega_l(z) = -\left[ K_{ll} \sum_0^{z-n_s} e^{-i/\zeta} + K_{ls} \sum_{z-n_s}^{z} e^{-i/\zeta} + K_{lS} \sum_{z}^{z+n_s} e^{-i/\zeta} \right] + K_{ll} \sum_0^{n_l+n_s-z} e^{-i/\zeta} \qquad \textbf{(I)}$$

where the first term corresponds to the total interaction between the z-layer and all the underneath layers (liquid for $0<i<z-n_s$, solid for $z-n_s<i<z$ and substrate for $z<i<z+n_s$), the constants $K_{\alpha\beta}$ describing the chemical interaction between materials $\alpha$ and $\beta$ ($\alpha,\beta$=S,s or l). The last term corresponds to the interaction between the z-layer and the upper liquid layers. Since the attraction by the upper layers and the underneath layers are in opposite directions, these two terms have not the same sign. In other words there is some compensation in between the upper and the underneath attraction.

This interaction field has to be compared to the interaction field $\omega_{l,0}(z)$ felt by a liquid layer at z in a continuous liquid environment having the same geometry ($n_s+n_l+n_s$ layers). This field can simply be obtained by writing $K_{\alpha\beta}=K_{ll}$ in formula (I) so that there is:

$$\omega_{l,0}(z) = -K_{ll} \sum_{0}^{z+n_s} e^{-i/\zeta} + K_{ll} \sum_{0}^{n_l+n_s-z} e^{-i/\zeta} \qquad \text{(II)}$$

The excess field felt by the liquid z-layer in the composite slab (S/s/l) in respect to the same layer in a continuous liquid material ($n_S+n_s+n_l$ layers) can thus be written $\Omega_l(z) = \omega_l(z) - \omega_{l,0}(z)$. Calculating thus the summation and using the same procedure as in appendix III to identify the material constant $K_{\alpha\beta}$, that gives:

$$K_{\alpha\alpha}/\left(1-e^{-1/\zeta}\right) = 2\gamma_\alpha^\infty \qquad \text{and} \qquad K_{\alpha\beta}/\left(1-e^{-1/\zeta}\right) = -\beta_{\alpha/\beta}^\infty \qquad \text{(III)}$$

there is:

$$\Omega_l(z) = e^{-z/\zeta}\left[2\gamma_l\left(e^{n_s/\zeta} - e^{-n_S/\zeta}\right) - \beta_{l/s}\left(1-e^{-n_S/\zeta}\right) - \beta_{l/S}\left(e^{-n_S/\zeta}-1\right)\right] \quad ; n_s < z < n_s + n_l \qquad \text{(IV)}$$

Where for the sake of simplicity we omit the $\infty$ subscripts.

- For a solid layer ($z<n_s$) the summation of the interactions between the z-layer and the whole of the material reads:

$$\omega_s(z) = -\left[K_{ss}\sum_{0}^{z} e^{-i/\zeta} + K_{sS}\sum_{z}^{z+n_s} e^{-i/\zeta}\right] + \left[K_{ss}\sum_{0}^{n_s-z} e^{-i/\zeta} + K_{sl}\sum_{n_s-z}^{n_s+n_l-z} e^{-i/\zeta}\right] \qquad \text{(V)}$$

where the first term again corresponds to the interaction with all the underneath layers ( s for $0<i<z$, S for $z<i<z+n_S$). It opposes to the second term which corresponds to the interaction with all the upper layers (solid for $0<i<n_s-z$, liquid for $n_s-z<i<n_s+n_l-z$). The field felt by the same z-layer in a continuous solid material having $n_S+n_s+n_l$ layers again is simply obtained by substituting $K_{ss}$ for $K_{\alpha\beta}$ in (V) and thus reads:

$$\omega_{s,0}(z) = -K_{ss}\sum_{0}^{n_s+n_l-z} e^{-i/\zeta} + K_{ss}\sum_{0}^{z+n_s} e^{-i/\zeta} \qquad \text{(VI)}$$

The excess field felt by the solid z-layer in the composite slab (S/s/l) in respect to the same layer in a continuous solid material ($n_S+n_s+n_l$ layers) can thus be written $\Omega_s(z) = \omega_s(z) - \omega_{s,0}(z)$ which gives after summation and identification of the material constant:

$$\begin{cases} \Omega_s(z) = (2\gamma_s - \beta_{s/S})e^{-z/\zeta}\left(1-e^{-n_S/\zeta}\right) - (2\gamma_s - \beta_{s/l})e^{-(n_s-z)/\zeta}\left(1-e^{-n_l/\zeta}\right) \\ \Omega_S(z) = (\beta_{l/S} - \beta_{s/S})e^{-(|z|+n_s)/\zeta} + (2\gamma_S - \beta_{l/S})e^{-(|z|+n_s+n_l)/\zeta} - (2\gamma_S - \beta_{l/S})e^{-|z|/\zeta} \end{cases} \qquad \text{(VII)}$$

# Appendix VIII
## List of principal symbols

$A$ : surface area

$a,b$: crystallographic parameter of materials A and B (supposed to be cubic)

$\beta_{i/j}$ : adhesion energy between materials i and j

$c_p$ : heat capacity at constant pressure

$\chi_l$ : compressibility of the liquid

$d$: interplane parameter

$\Delta S_m$ : latent melting entropy

$E$: Young modulus

$\varepsilon_{ij}$ : strain tensor component

$\Phi$ : l/s wetting factor

$\tilde{\Phi}$ : s/l wetting factor

G: Gibbs energy

$\gamma_i$ : surface energy of material i (put in vacuum)

$\gamma_{ij}$ : interfacial energy between materials i and j

$K_{ij}$ : chemical interaction between materials i and j

$m$: misfit

$\mu^f$ : chemical potential of the fluid

$n_l$ : number of liquid layers of the deposited material

$n_s$ : number of solid layers of the deposited material

$n$ : total number of layers of the deposited material

$N_{av}$ : Avogadro number

$\nu$ : Poisson ratio

$\omega_i(z)$ :interaction field felt by a layer of material i located at level z

$\Omega_i(z)$ : excess  stress field felt by a layer of material i at level z

$P$: hydrostatic pressure

$S$: entropy

$s_i$ : surface stress of material i (put in vacuum)

$s_{ij}$ : interfacial stress between materials i and j

$\sigma_{ij}$ : stress tensor component

$T_m$ : usual melting temperature

$T'_m$ : melting temperature of homogeneously strained solid

$T_s$ : temperature below which $n_l$ =0

$T_l$ : temperature beyond which $n_s$ =0

$U$ : internal energy

$V^i$ : molar volume of material i

$Y_s$ : linear compressibility of the solid

$Z$ : bulk coordination

$z$ : surface coordination

$\zeta_i$ : screening parameter for material i

# Figure captions

**Figure 1:** Schematic drawing of surface induced melting (a) of a semi-infinite solid s and (b) of the surface induced freezing of the semi-infinite liquid l. The conditions are given in terms of temperature T in respect to the melting temperature $T_m$, of surface energy $\gamma_i$ and adhesion energy $\beta_{ij}$ so that in (a) the liquid wets the solid s or the reverse s wets the liquid l in (b).

**Figure 2:** Schematic drawing of surface induced melting in (a) of a solid thin film s supported by a thick substrate S whose melting temperature is $T_S$. Similar case of surface induced freezing in (b) of a thin liquid film l supported by S. In the first line are given wetting conditions of l/s (or (s/l) as in figure 1. In the second line are written the stability conditions for having a uniform film, s/S when preparing these films. When the deposits s is molten this second relation reads $2\gamma_l - \beta_{ls} < 0$.

**Figure 3:** Stable premelting curves of system fig 2a with $n_l$ the number of liquid layers versus T. $T_m^{'}$ the bulk melting point of the coherent stressed solid film of thickness $n_s = n - n_l$. Four cases with the same wetting parameter $\Phi < 0$. **(1)** Premelting of a thick solid ($n_s = \infty$) reaching asymptotically $T_m^{'}$. **(2)** Thin solid of n layers. $\Gamma = 0$: same premelting curve as **(1)** but ending in $n_l = n$. **(3)** Case $\Gamma < 0$ n-finite, melting astride $T_m^{'}$ with its overheating zone $T > T_m^{'}$ **(4)** $\Gamma > 0$, n-finite, premelting going over continuously at $T^* < T_m^{'}$, $n_l \approx n/2$ in first order premelting. All curves have a common leading edge $e^{-n_l/\xi}$ at $T < T_m^{'}$.

**Figure 4:** System of fig. 2a where schematically are acting thermodynamic forces (arrows) on the liquid/solid interface s/l. The interface l/v due to $\Phi < 0$ acts similarly in both cases a) and b) to increase the amount of liquid (premelting). However interface s/S acts against when $\Gamma < 0$ in (a) or as in (b) $\Gamma > 0$ acts with.

**Figure 5:** The astride melting case $\Phi < 0$, $\Gamma < 0$, n=10. In (a) $\Phi = -50$, $\Gamma = -10$, -50,-100 erg.cm$^{-2}$. In (b) $\Gamma = -10$, $\Phi = 0$,-10,-50, -100 erg.cm$^{-2}$. Wetting $\Phi$ is sensitive to premelting, insensitive to overheating. $\Gamma$ is insensitive to premelting but sensitive to overheating.

**Figure 6:** The boosted premelting. When $\Phi<0$, $\Gamma>0$, n=10 the stable branches (full lines) are only sensitive to $\Phi$ (see (a)). The unstable branches (dotted) are only sensitive to $\Gamma$ (see (b)). At the end-points n*, the intersection of the curves with a vertical line, starts the first order premelting.

**Figure 7:** The melting curve of CH4/MgO. Continuous curve: the calculated (12) one $\Phi=-3.7$, $\Gamma=-10.6$ erg cm$^{-2}$ (the misfit dependence with temperature has been taken into account but in this specific case, due to its weakness, is negligible). From $T_s$ to T'$_m$ (90 K) this curve fits well the experimental measurements [20]. They are not reported here. However the measured values at T>T'$_m$ typical for astride melting are given with their error bars. The dotted curve is that one (12') corresponding to van der Waals interaction.

**Figure 8:** In plane strain $\varepsilon_{eq}$ of solid film s of system fig 2a, when the epitaxy of s/S is incoherent and perfectly glissile, versus relative number of solid layers 0<n$_s$/n<1. The total number of layers n=n$_s$+n$_l$ passes from n=10 to n=100. The interfacial stresses s$_{ij}$ are taken positive so that the strain is negative.

**Figure 9:** Melting curves (full) of a glissile epitaxy of n=15 layers with the same elastic strain as in figure 8. For comparison (dotted) the system without strain. In (a) astride melting $\Phi=-30$, $\Gamma=-50$ erg cm$^{-2}$; in (b) boosted premelting $\Phi=-30$, $\Gamma=50$ erg cm$^{-2}$. In this examples when $(s_s+s_{sS})\Phi'<0$ the premelting zone is repressed.

**Figure 10:** Kr/graphite
a) The peak values of the experimental curves of [22] versus n the number of Kr layers fit an exponential law (stars). They fit too a power law (dots and upper abscissae) but with a power 1.4 instead of 4 meaningful for van der Waals interactions.
b) The experimental peak shifts of [22] satisfy a $\left(1-e^{-n/\zeta}\right)\!/n$ square law typical for a glissile epitaxy.

**Figure 11:** a) Experimental excess $C(t)$ heat capacity [23] with reduced temperature $t=\left(T-T_m\right)/T_m$

b) Calculated one according to the formulation for glissile epitaxy. For a) and b) the curves correspond to  n=11.8; 10.3; 8.7; 7.2  the different total Kr coverages. C(t) in J/K units.

**Figure 12:** Solid (s) of 4 layers (0<z<4) adheres on substrate (S )on the left. Deposit s is covered by 4 layers of its melt (l) (4<z<8) which wets it perfectly. The field induced by the substrate S due to $2\gamma_s - \beta_{sS}$ <0 compresses $\varepsilon_{zz}$<0 mostly the nearest layers of s on S. The liquid feels less this field but that of s so that since $2\gamma_s - \beta_{sl}$<0  this liquid is also inhomogeneously compressed by this field. The data used are those of $CH_4$/MgO at 90K.

**Figure 13:** The melting curve of a coherently bond epitaxial layer is not appreciably changed by considering the inhomogeneous z-strain due to the interfacial fields excepted for the last melting layers. The curves represent the system $CH_4$/MgO: full curve reference, dotted curve with inhomogeneous effect. The left hand insert gives the total elastic energy (full curve) and misfit energy (dotted) as a function of solid layers $n_s$ . The remainder $n-n_s=n_l$ is the number of liquid layers. Here n=8. The right hand insert gives the melting point distortion versus $n_s$ (full curve), dashed the shift due to the misfit.

| Z | 13Al | 23V | 25Mn | 26Fe | 27Co | 29Cu | 30Zn | 31Ga | 45Rh | 46Pd | 48Cd | 49In | 50Sn | 78Pt | 79Au | Tl81 | 82Pb | 83Bi |
|---|---|---|---|---|---|---|---|---|---|---|---|---|---|---|---|---|---|---|
| $\Phi$ | -13 | -5 | -14 | -50 | -22 | -19 | -6 | -21 | -38 | -26 | -26 | -30 | -18 | -29 | -33 | -16 | -22 | -47 |
| $\widetilde{\Phi}$ | 321 | 703 | 380 | 702 | 712 | 545 | 244 | 137 | 806 | 630 | 188 | 126 | 150 | 674 | 433 | 148 | 146 | 195 |

**Table I :** Wetting factors $\Phi$ and $\widetilde{\Phi}$ of elements $Z$ (from [11]). The meaning of $\Phi$ (formula (9)) is $\Phi = 2\gamma_l - \beta_{sl} = \gamma_l + \gamma_{ls} - \gamma_s$ and $\widetilde{\Phi}$ is obtained from $\Phi$ by interchanging l and s.

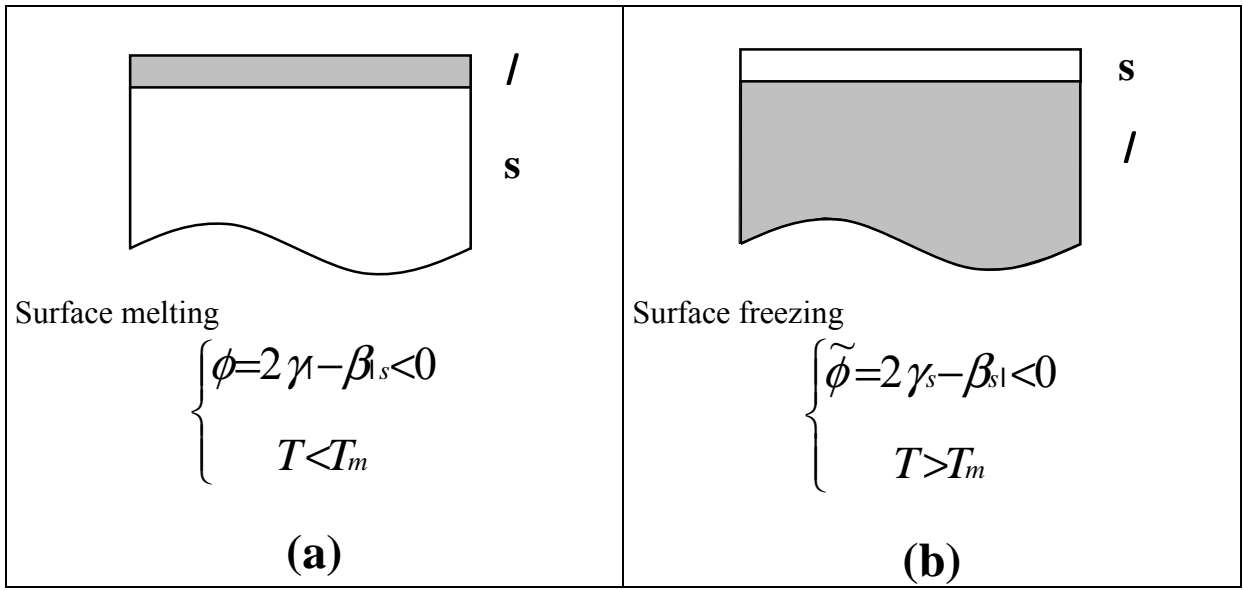

Fig 1

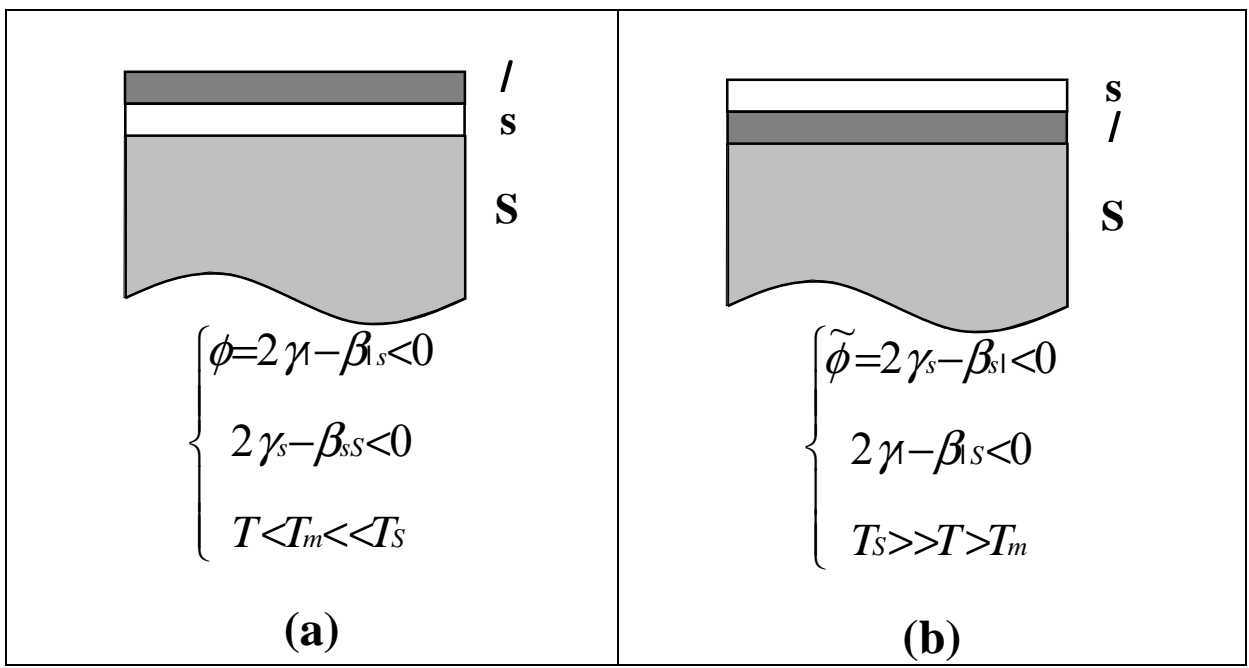

Fig 2

**Fig 3:**

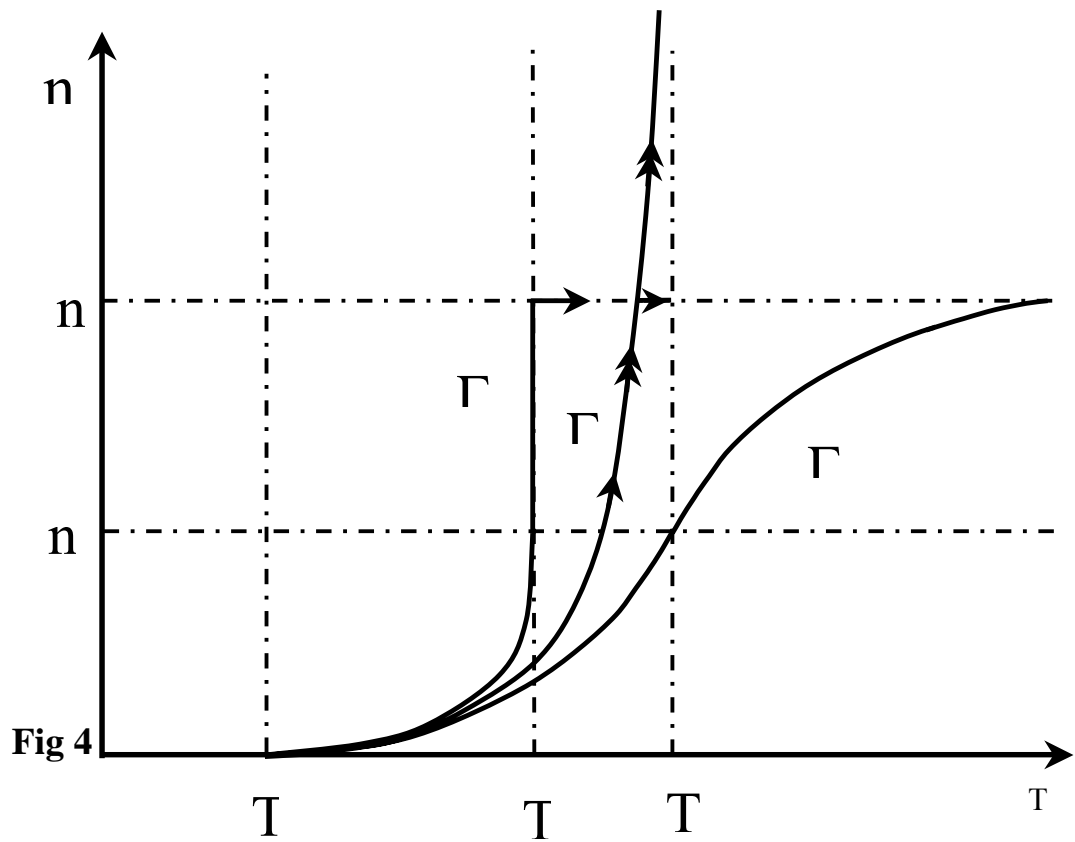

**Fig 4**

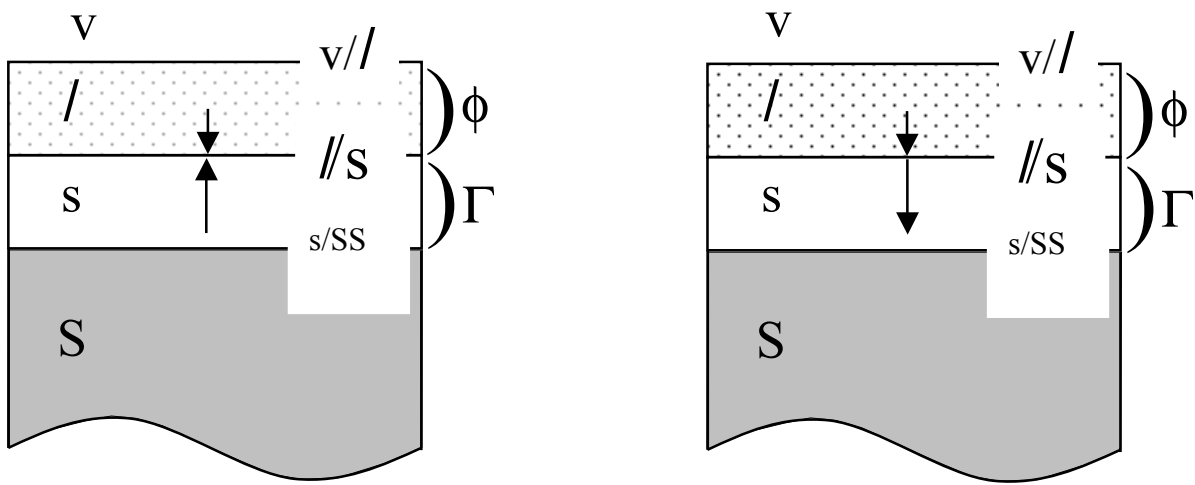

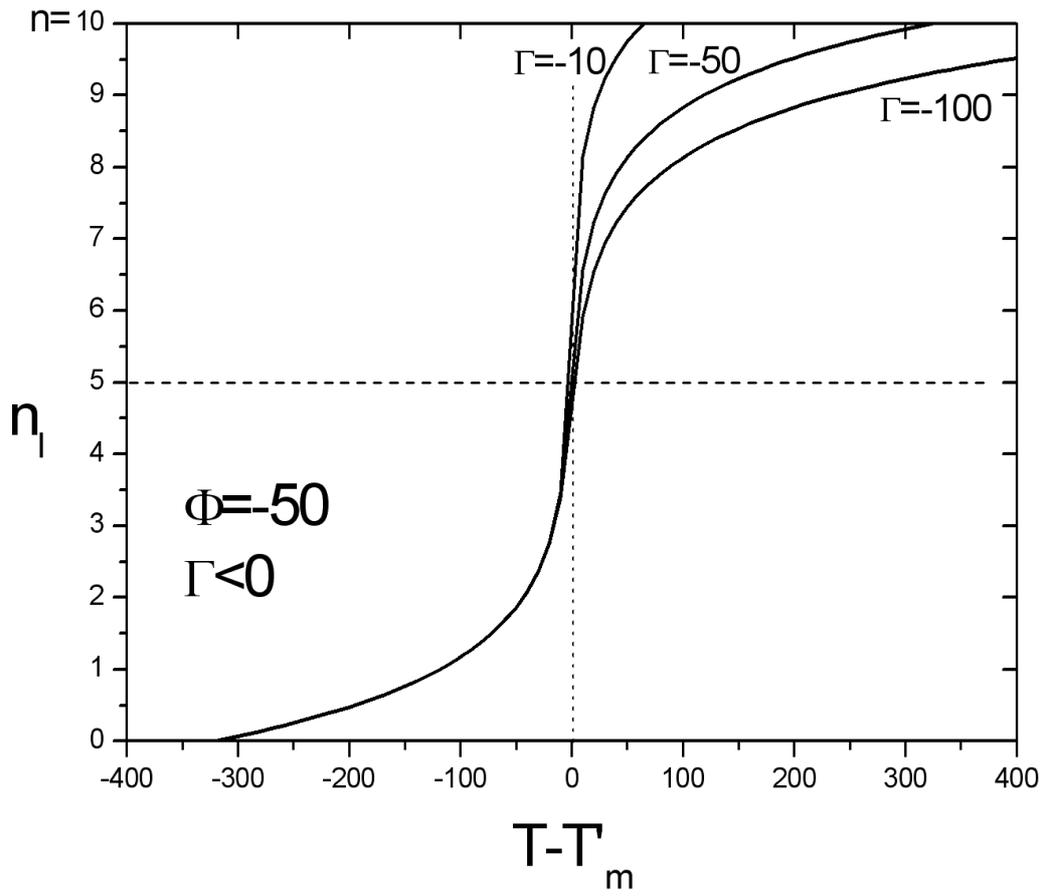

Fig 5a

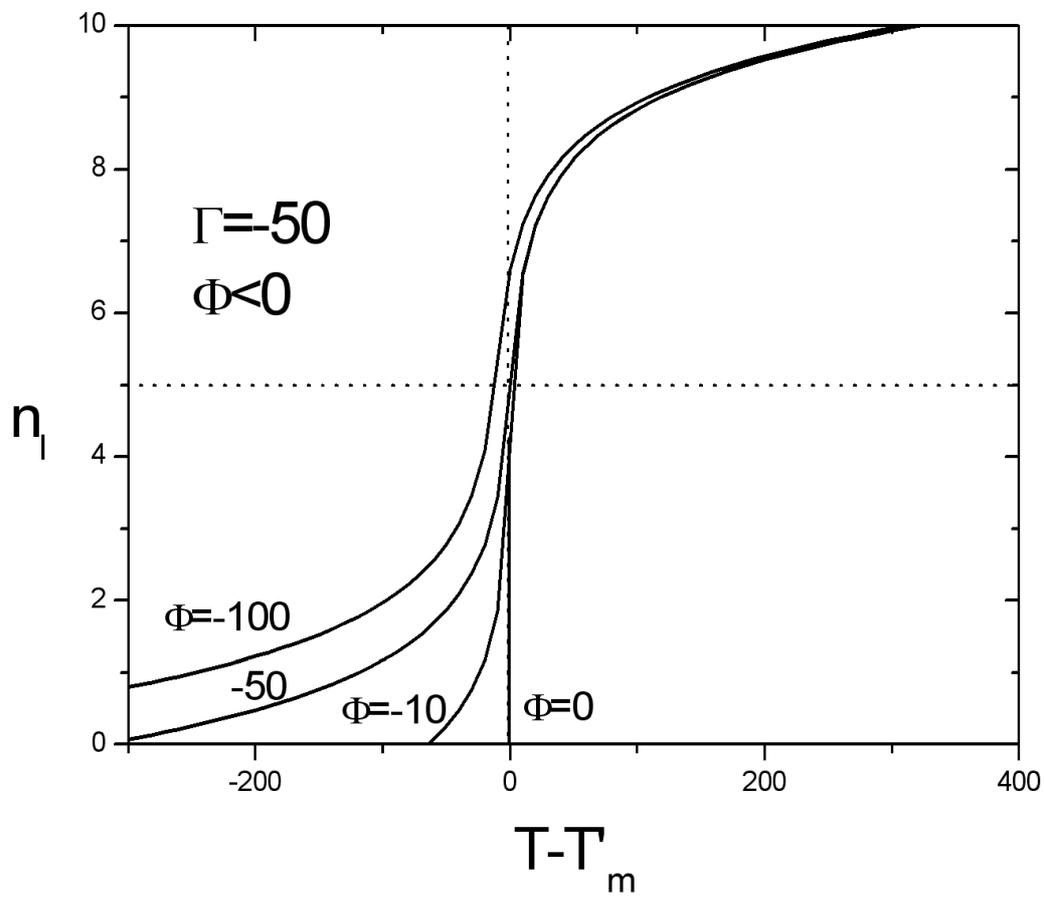

Fig 5b

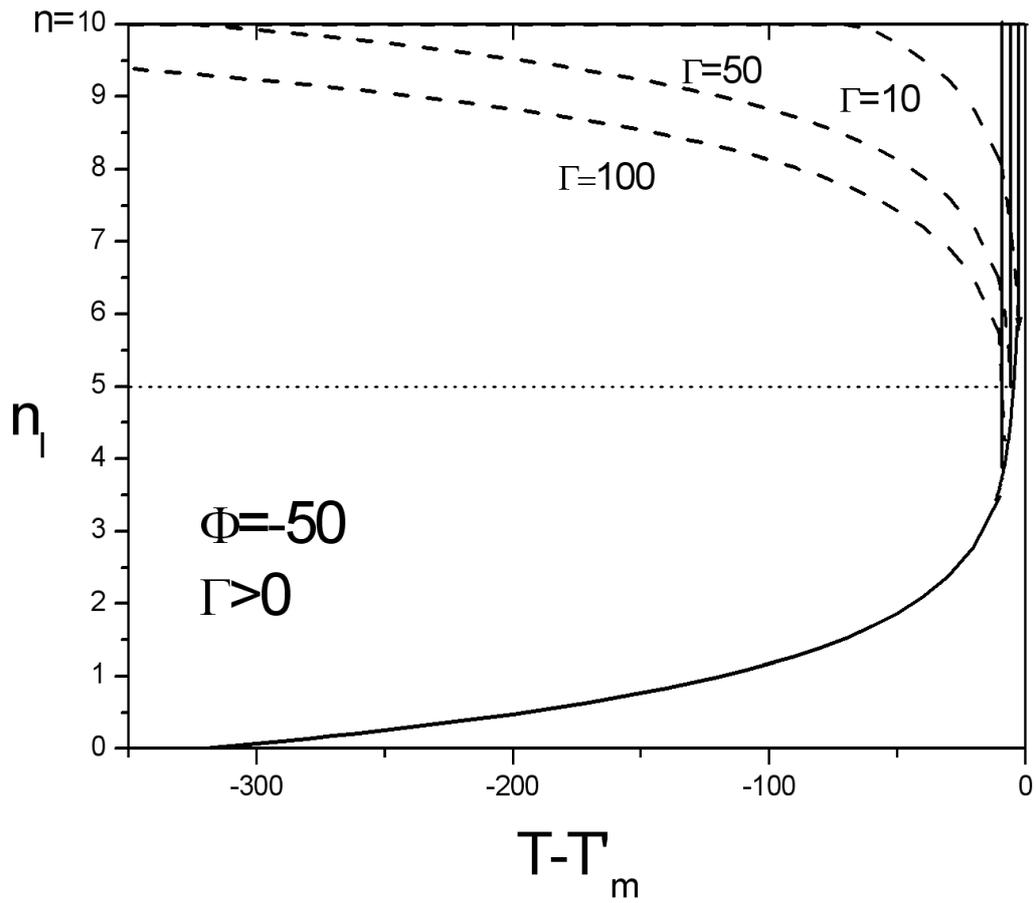

Fig 6a

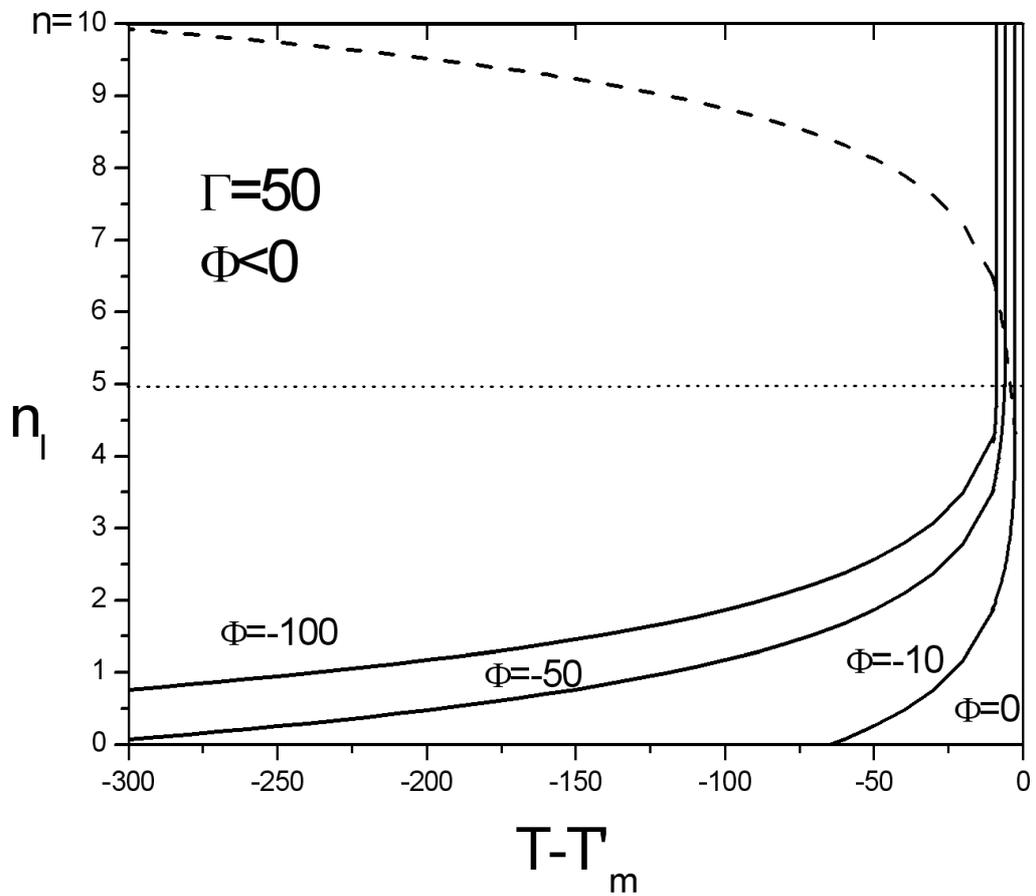

Fig6b

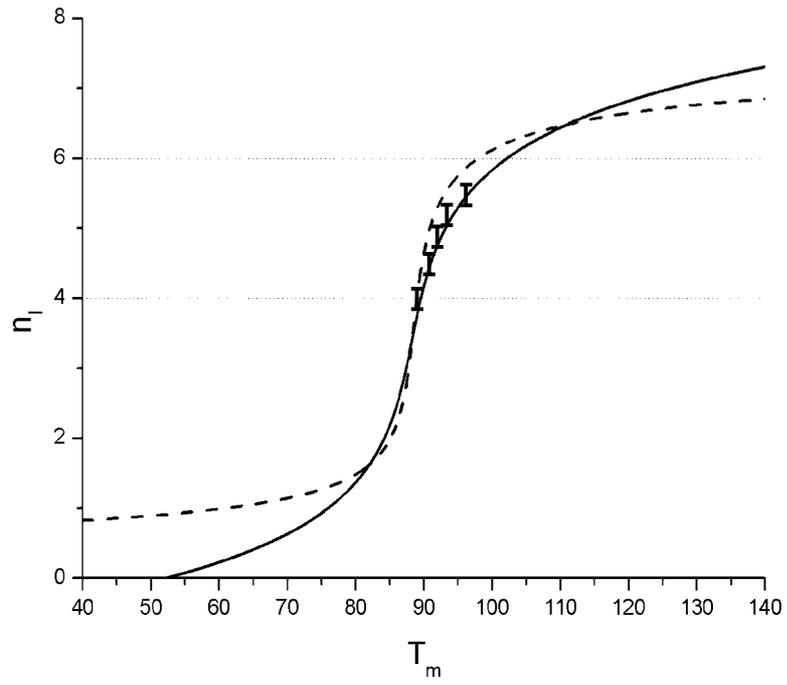

Fig 7

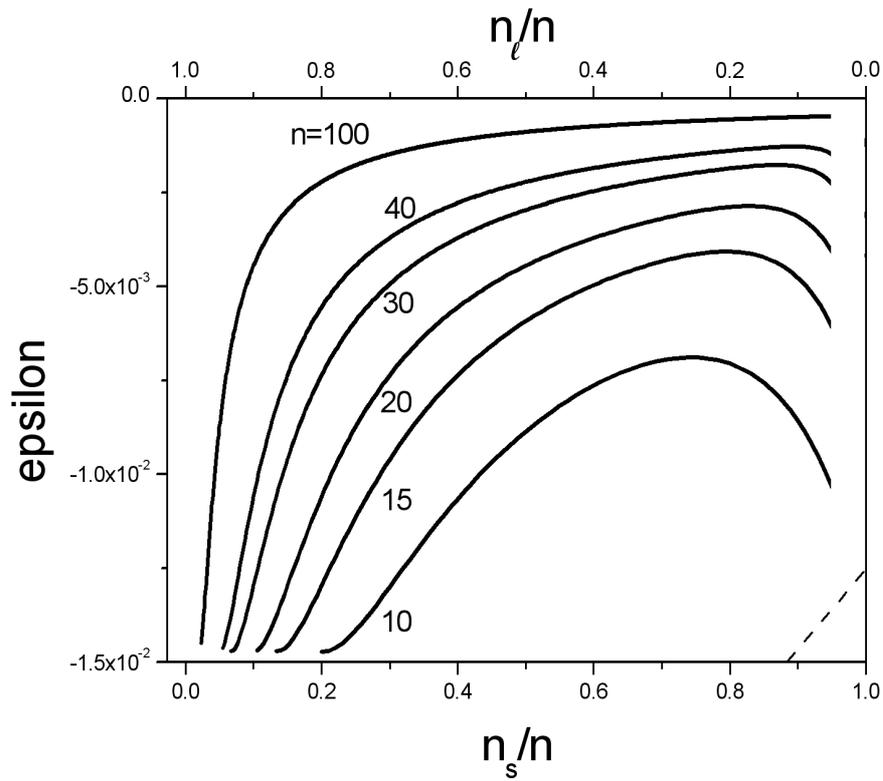

Fig 8

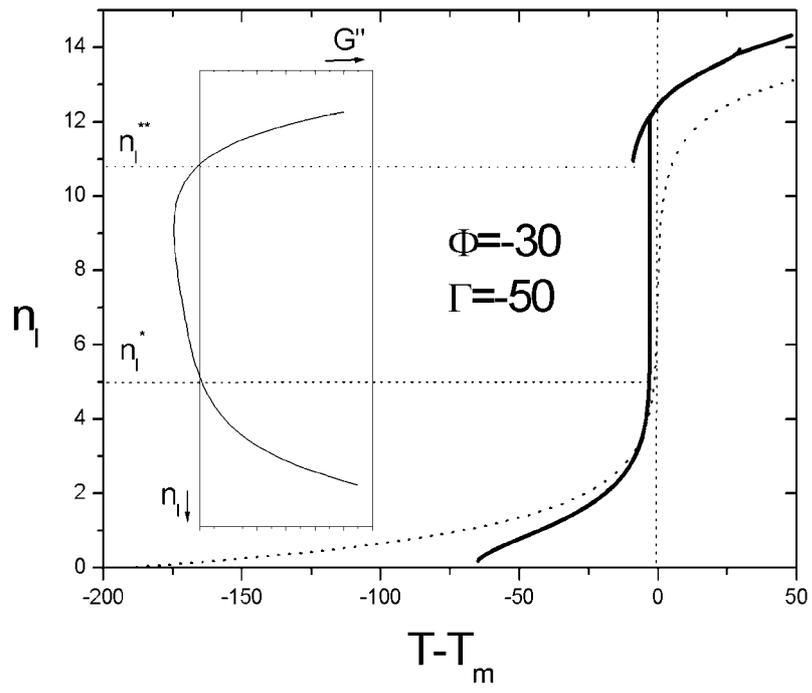

Φ=-30
Γ=-50

Fig 9a

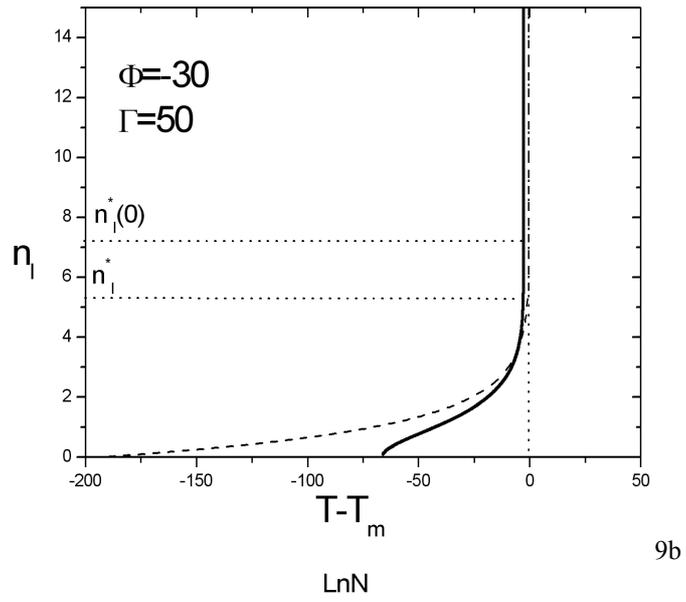

9b

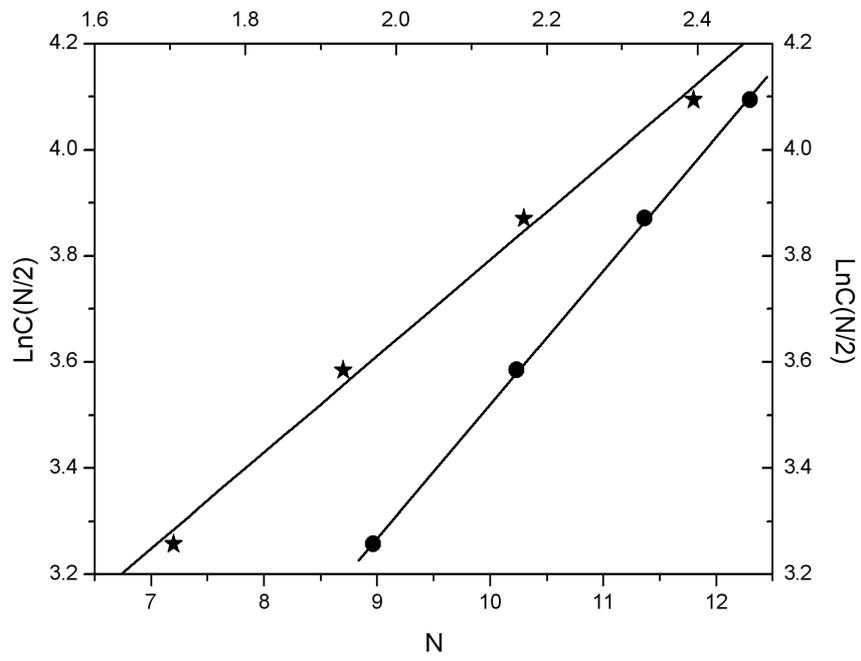

Fig 10a

Fig 10b

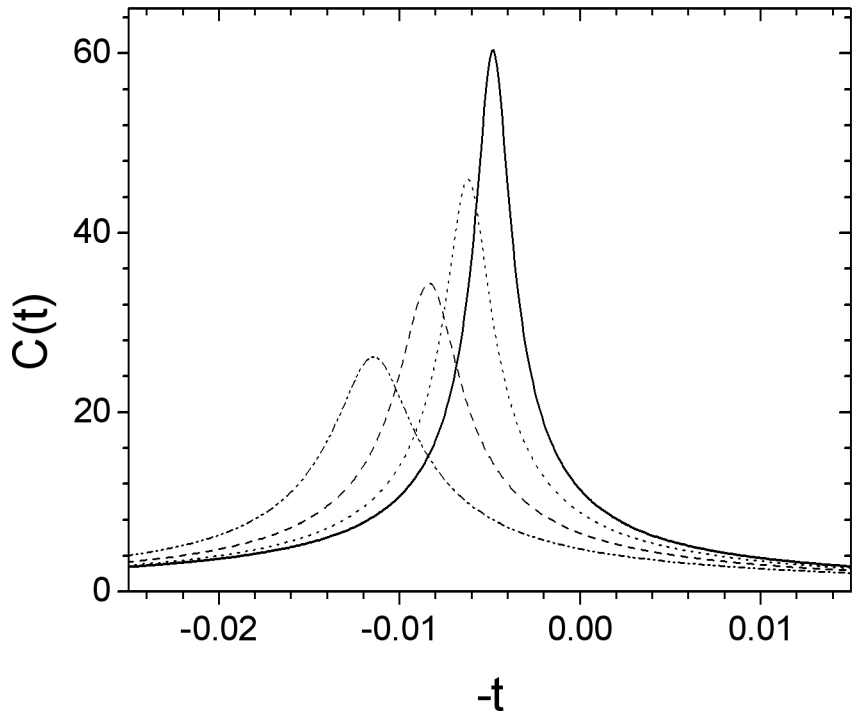

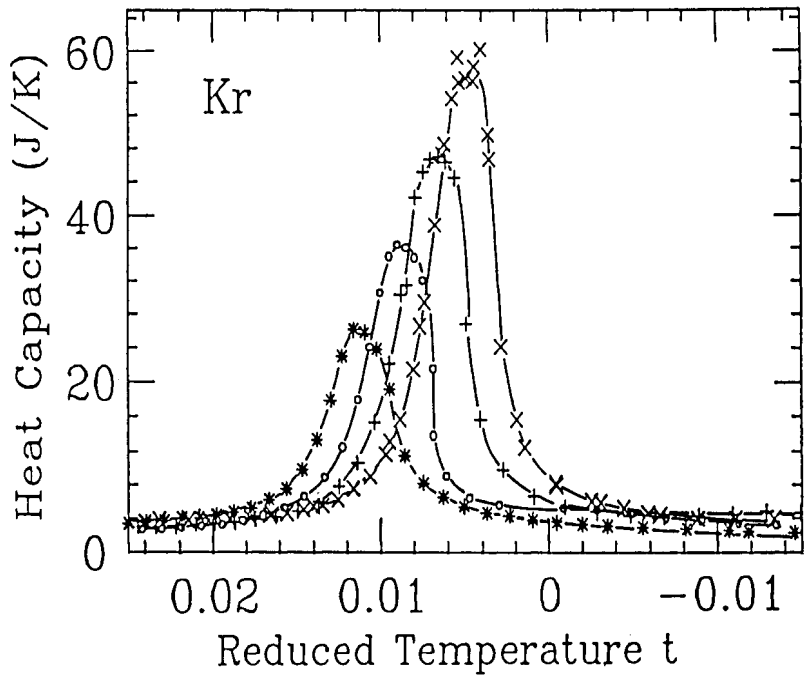

Fig 11

Fig 13

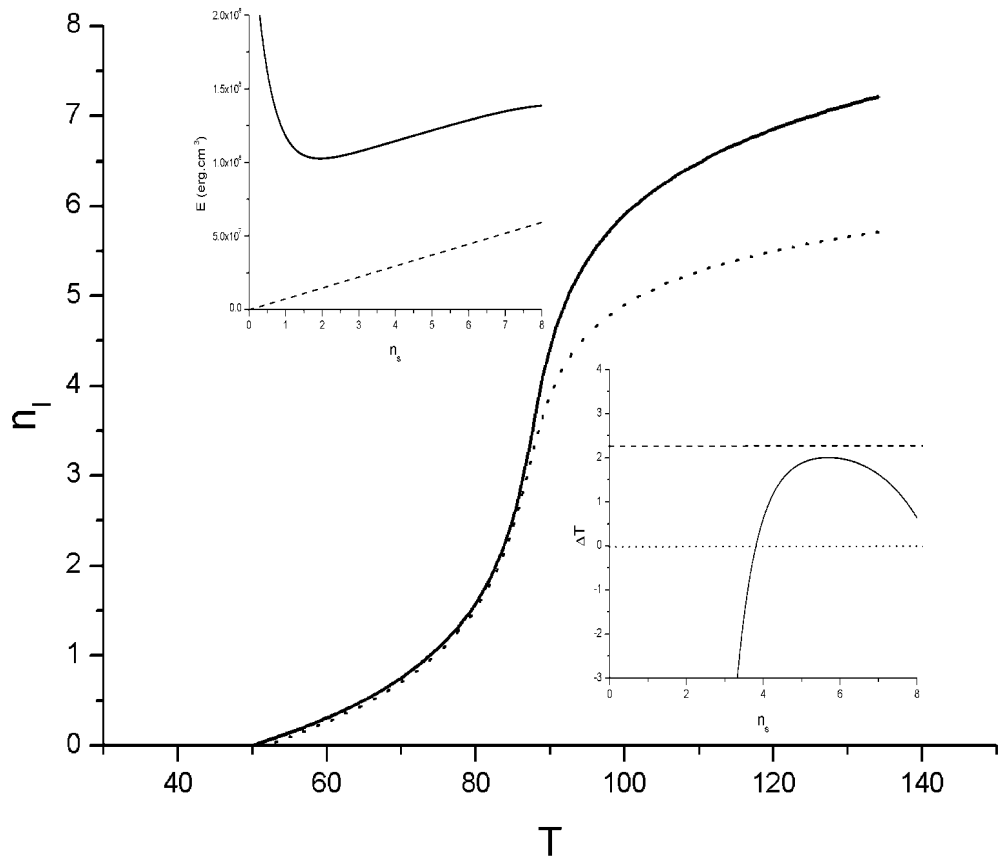